\documentclass[a4paper,12pt]{article}
\usepackage{amsmath}\usepackage{amssymb}
\usepackage{latexsym}\usepackage{cite}

\usepackage{amsmath}
\usepackage[dvips]{graphicx}
\usepackage{pstricks}
\usepackage{color}
\usepackage{bm}

\usepackage{graphicx}
\usepackage{caption}
\usepackage{varioref}
\usepackage{placeins}
\usepackage{float}
\usepackage{amssymb}
\usepackage{amsmath}
\usepackage{graphicx}
\usepackage{verbatim}
\usepackage{color}
\usepackage{subfigure}
\usepackage{hyperref}
\usepackage[final]{pdfpages}
\usepackage{afterpage}
\usepackage{makecell}
\usepackage{multirow}
\usepackage{caption}

\definecolor{Red}{rgb}{1.00, 0.00, 0.00}
\definecolor{Green}{rgb}{0.00, 1.00, 0.00}
\definecolor{Blue}{rgb}{0.00, 0.00, 1.00}
\definecolor{Cyan}{rgb}{0.00, 1.00, 1.00}
\definecolor{Mymagenta}{rgb}{0.3, 0.00, 1.00}%
\definecolor{Magenta}{rgb}{1.00, 0.00, 1.00}
\definecolor{DeepSkyBlue}{rgb}{0.00, 0.75, 1.00}
\definecolor{DarkGreen}{rgb}{0.00, 0.39, 0.00}
\definecolor{SpringGreen}{rgb}{0.00, 1.00, 0.50}
\definecolor{Mygreen}{rgb}{0.00, 0.72, 0.00}
\definecolor{DarkOrange}{rgb}{1.00, 0.55, 0.00}
\definecolor{OrangeRed}{rgb}{1.00, 0.27, 0.00}
\definecolor{DeepPink}{rgb}{1.00, 0.08, 0.57}
\definecolor{DarkViolet}{rgb}{0.58, 0.00, 0.82}
\definecolor{SaddleBrown}{rgb}{0.57, 0.27, 0.07}
\definecolor{Black}{rgb}{1.00, 1.00, 1.00}
\definecolor{Ablue}{rgb}{0.10, 0.1, 1.00}

\newcommand{\tr}{\,\textup{tr}}
\newcommand{\ka}{\kappa}

\def\beq{\begin{equation}}
\def\eeq{\end{equation}}
\def\la{\label}
\def\beqs{\begin{equation*}}
\def\eeqs{\end{equation*}}

\def\beq{\begin{equation}}
\def\eeq{\end{equation}}
\def\beqr{\begin{eqnarray}}
\def\eeqr{\end{eqnarray}}

\def\al{\alpha}
\def\bt{\beta}
\def\Ga{\Gamma}

\def\de{\delta}
\def\De{\Delta}

\def\ka{\kappa}
\def\si{\sigma}

\def\te{\theta}

\def\lam{\lambda}

\def\om{\omega}
\def\ep{\epsilon}

\def\sq{\sqrt}

\def\l{\left (}
\def\r{\right )}
\def\lq{\left [}
\def\rq{\right ]}
\def\fr{\frac}
\def\la{\label}
\def\hs{\hspace}
\def\vs{\vspace}

\def\ov{\overline}
\def\tl{\tilde}
\def\tm{\times}

\def\lrar{\leftrightarrow}

\def\bea{\begin{eqnarray}}
\def\eea{\end{eqnarray}}

\allowdisplaybreaks
\begin{document}
\begin{flushright}
October 27, 2017 \\
CERN-TH-2017-202
\end{flushright}

\vs{1cm}

\begin{center}
{\Large\bf

 Texture Zero Neutrino Models and Their Connection with Resonant Leptogenesis

 %
 %
 }

\end{center}

\vspace{0.5cm}
\begin{center}
{\large
~Avtandil Achelashvili\footnote{E-mail: avtandil.achelashvili.1@iliauni.edu.ge},~ and
~Zurab Tavartkiladze\footnote{E-mail: zurab.tavartkiladze@gmail.com}
}
\vspace{0.5cm}

{\em Center for Elementary Particle Physics, ITP, Ilia State University, 0162 Tbilisi, Georgia}
\end{center}

\vspace{0.6cm}

\begin{abstract}

Within the low scale resonant leptogenesis scenario, the cosmological CP asymmetry may arise
by radiative corrections through the charged lepton Yukawa couplings. While in some cases, as one expects, decisive
role is played by the $\lambda_{\tau }$ coupling, we show that in specific neutrino textures only by inclusion of the
$\lambda_{\mu }$ the cosmological CP violation is generated at 1-loop level.

With the purpose to relate the cosmological CP violation to the leptonic CP phase $\delta $, we consider an extension of MSSM with two right
handed neutrinos (RHN), which are degenerate in mass at high scales. Together with this, we first consider two texture zero
$3\times2$  Dirac Yukawa matrices of neutrinos. These via see-saw generated neutrino mass matrices augmented by single
 $\Delta L=2$ dimension five ($\rm d=5$) operator  give predictive neutrino sectors with calculable CP asymmetries.
 The latter is generated through $\lambda_{\mu , \tau }$ coupling(s) at 1-loop level. Detailed analysis of the leptogenesis
 is performed. We also revise some one texture zero Dirac Yukawa matrices, considered earlier, and show that addition of a
 single $\Delta L=2$, $\rm d=5$ entry in the neutrino mass matrices, together with newly computed 1-loop corrections to the
 CP asymmetries, give nice accommodation of the neutrino sector and desirable amount of the baryon asymmetry via the resonant leptogenesis
 even for rather low RHN masses($\sim $few TeV -- $10^7$~GeV).

\end{abstract}

\hspace{0.4cm}{\it Keywords:}~CP violation; Resonant Leptogenesis; Neutrino mass and mixing; Renormalization.

\hspace{0.4cm}PACS numbers:~11.30.Er, 98.80.Cq, 14.60.Pq, 11.10.Gh.

\section{Introduction}
\la{intro}

Problem of neutrino masses and generation of the baryon asymmetry of the Universe, together with the
dark matter problem and naturalness issues, call for some reasonable extension(s) of the Standard Model (SM).
Perhaps simplest and most elegant simultaneous resolution of the first two puzzles is by the SM extension
with the right handed neutrinos (RHN). This, by  the $\Delta L=2$ lepton number violating interactions generates the neutrino masses
via celebrated see-saw mechanism \cite{seesaw},\cite{Schechter:1980gr}, accommodating the atmospheric and solar neutrino data \cite{recent-nu-data}, and gives an elegant possibility
for the baryogenesis through the thermal leptogenesis \cite{Fukugita:1986hr} (for reviews see Refs. \cite{{Giudice:2003jh}, {Buchmuller:2004nz}, {Davidson:2008bu}}).

Motivated by these, we consider the minimal supersymmetric standard model (MSSM)\footnote{This setup with the SUSY scale $M_S\sim$ few TeV
guarantees the natural stability of the EW scale.}
augmented by two degenerate RHNs. Note that the degeneracy in the RHN mass spectrum offers  an elegant possibility of resonant leptogenesis \cite{{Flanz:1996fb}, {Pilaftsis:1997jf}, {Pilaftsis:2003gt}}
(see \cite{{Blanchet:2012bk}, {Dev:2014laa},{Dev:2015wpa}, {Pilaftsis:2015bja}, {Achelashvili:2016trx}} for recent discussions on resonant leptogenesis). This framework, as it was shown in \cite{{Babu:2007zm}, {Babu:2008kp}, {Achelashvili:2016trx}}, with specific forms of the Yukawa couplings, allows to have highly predictive model. In particular, in \cite{Achelashvili:2016nkr} all possible two texture zero $3\tm 2$ Dirac type neutrino Yukawa couplings have been considered. Those, via see-saw generated neutrino mass matrices augmented by a single
${\rm d=5}$, $\Delta L=2$ operator, gave consistent neutrino scenarios. As it was shown, all experimentally viable cases allowed to calculate the cosmological CP violation in terms
of a single known (from the model) leptonic phase $\delta $.\footnote{The approach
with texture zeros has been
put forward in \cite{Frampton:2002qc}, which successfully  relates the phase $\delta $ with the cosmological CP asymmetry \cite{{Frampton:2002qc}, {Ibarra:2003up}, {Shafi:2006nt}, {Branco:2006hz}}, \cite{Babu:2007zm}, \cite{Babu:2008kp},\cite{{Meroni:2012ze},{Harigaya:2012bw}, {Ge:2010js}}, \cite{Achelashvili:2016nkr}, \cite{Achelashvili:2016trx}.} In the subsequent work \cite{Achelashvili:2016trx}, the quantum corrections, primarily due to the
$\lambda_{\tau }$ Yukawa coupling, have been investigated and, confirming earlier claim of Refs. \cite{early-tau-effect}, it was shown that the cosmological CP asymmetry arises at 1-loop order.\footnote{Studies of \cite{Babu:2008kp} included only $\lambda_{\tau}$'s 2-loop effects in the RG of the RHN mass matrix, which
give parametrically more suppressed cosmological CP violation in comparison with those evaluated in \cite{Achelashvili:2016trx}.}
Demonstrated on a specific fully consistent neutrino model \cite{Achelashvili:2016trx}, this was shown to work well and opened wide prospect for the model building for the low scale resonant leptogenesis.

The goals of this work are following. First we give detailed and conscious derivation of the loop induced leptonic cosmological
CP violation showing the necessity of inclusion of the charged lepton Yukawa couplings. Proof includes analytical expressions and is extended
by inclusion of the $\lambda_{\mu }$ coupling which as it turns out in specific neutrino
scenarios is the only relevant source of the cosmological CP violation within considered scenarios with the RHN
masses $\stackrel{<}{_\sim }10^7$~GeV.
We apply obtained result to specific neutrino textures. While in Refs.
\cite{{Frampton:2002qc}, {Ibarra:2003up}, {Shafi:2006nt}, {Branco:2006hz}}, \cite{Babu:2007zm}, \cite{Babu:2008kp},\cite{{Meroni:2012ze},{Harigaya:2012bw}, {Ge:2010js}}
 the textures
relating the cosmological CP violation to the leptonic $\delta $ phase (being still undetermined from the construction) have been discussed,
in \cite{Achelashvili:2016nkr} we have proposed models, which not only give such relations, but also predict the values of the $\delta $
(the leptonic  Dirac phase)
 and $\rho_{1,2}$ (two leptonic Majorana phases) and consequently the cosmological CP violation.
From the   constructions   of \cite{Achelashvili:2016nkr} we consider
 viable neutrino models built by two texture zero $3\tm 2$ Yukawa coupling generated see-saw neutrino mass matrices augmented by the single
$\Delta L=2$, ${\rm d=5}$ operator.
For all these neutrino models, applying obtained all relevant corrections,
 we investigate the resonant leptogenesis process, which has not been performed before. Along with the  cases
where crucial is $\lambda_{\tau }$ coupling, we have ones  for which the leptonic asymmetry originates due to the
$\lambda_{\mu }$ Yukawa coupling. Such possibility has not been presented before in the literature.
We also revise textures of \cite{Babu:2008kp} and consider their improved versions by addition of single ${\rm d=5}$ entry to the neutrino mass matrix, making them consistent  and  also viable for the baryogenesis.
 The details of the  calculation of the contribution to the leptonic asymmetry from the right handed sneutrino decays
are given as well. These include new corrections corresponding to the muon lepton soft SUSY breaking terms.  Also, refined and more
accurate expressions for the  decay widths and  absorptive parts, relevant for the CP asymmetries, are used.

Although in this work we are using the results of the loop induced cosmological CP violation (summarized in section \ref{RG-CP} and in
Appendixes \ref{app-RG},  \ref{app-scalar-asym}) for specific
texture zero models, the application can be extended to any model with two (quasi) degenerate RHNs.

The paper is organized as follows.
 In section \ref{RG-CP}, after defining the  setup with two degenerate RHNs, we give details of the calculation of the loop induced cosmological CP violation.
Mainly we follow the method of Ref. \cite{Achelashvili:2016trx}
 proving inevitable emergence of the cosmological CP violation via
 charged lepton Yukawas at 1-loop level,  confirming earlier result of \cite{early-tau-effect} (which took into account $\lambda_{\tau }$ coupling). We also include the contribution due to the $\lam_{\mu }$ which has not been considered before.
In section \ref{RG-CP-1}, first we list all
 possible two texture zero $3\times 2$ Yukawa matrices, considered in \cite{Achelashvili:2016nkr}.
The see-saw neutrino mass matrices, obtained from these Yukawa textures, are augmented by the addition of single $\Delta L=2$, $\rm d=5$ mass terms to certain zero entries.
This makes the list of the phenomenologically viable and predictive neutrino mass matrices.
From them we pick up those which involve complexities and have potential for the CP asymmetry.
With the updated neutrino data, we give updated results of the corresponding neutrino models which are
highly predictive and determine cosmological CP violating phases in terms of the $\delta $ phase.
In section \ref{Resonant}, applying results of the previous sections we determine cosmological CP violation for each considered model
and use them for calculating of the baryon asymmetry. The latter is generated via resonant leptogenesis. We demonstrate that
successful scenarios are possible for the low RHN masses (in a range few TeV -- $10^7$ GeV).
In section \ref{improvement} we revise textures of Ref. \cite{Babu:2008kp} and make model improvements of the obtained neutrino mass matrices by adding
the single $\Delta L=2$, $\rm d=5$ mass terms to certain non-zero entries (in a spirit of Sect.\ref{RG-CP-1}).
This makes the neutrino scenarios compatible with the best fit values of the neutrino data \cite{recent-nu-data} and
also proves to blend well with the leptogenesis scenarios.
We stress that in the $P_4$ neutrino texture scenario (discussed in Sect. \ref{RG-CP-1}) and also in the texture ${\rm B_2}'$
(considered in Sect. \ref{improvement}), for successful leptogenesis to take place crucial role is played by the $\lambda_{\mu }$ Yukawa coupling
which via 1-loop correction generates sufficient amount of the cosmological CP asymmetry. Such possibility has not been
considered in the literature before. (The general expressions for the corresponding corrections are presented
in Sect.  \ref{RG-CP}).
Sect. \ref{discussions} includes discussion and outlook where we also summarize our results and highlight some prospects for a future work.
Appendix \ref{app-RG} includes some expressions, details related to the renormalization group (RG) studies and description of calculation procedures we are using.
In Appendix \ref{app-scalar-asym} the contribution to the net baryon asymmetry from the decays
of the scalar components (RHS) of the RHN superfields is considered in detail. These analyses also include new corrections due to $\lam_{\mu}$
and corresponding soft SUSY breaking trilinear $A_{\mu }$ coupling (besides $\lam_{\tau}$, $A_{\tau }$ and other relevant couplings).

\section{Loop Induced Calculable Cosmological CP Violation }
\la{RG-CP}
\numberwithin{equation}{section}

Before going to the calculations we first describe our setup. The framework is the MSSM augmented with two
right-handed neutrinos $N_{1}$ and $N_{2}$. This extension is enough to build consistent neutrino sector
accommodating the neutrino data \cite{recent-nu-data} and also to have a successful leptogenesis scenario.
The relevant lepton superpotential couplings are given by:
\begin{equation}
W_{lept}=l^{T}Y_{e}^{\rm diag}e^{c}h_{d}+l^{T}Y_{\nu}Nh_{u}-\frac{1}{2}N^{T}M_{N}N \la{r21},
\end{equation}
where $h_{d}$ and $h_{u}$ are down and up type MSSM Higgs doublet
superfields respectively and  $l^{T}=(l_{1}, l_{2}, l_{3})$, $e^{cT}=(e^{c}_{1}, e^{c}_{2}, e^{c}_{3})$, $N^T=(N_1, N_2)$.
 We  work in a basis in which the charged lepton Yukawa matrix is diagonal and real:
 \beq
 Y_{e}^{\rm diag}={\rm Diag}(\lambda_{e}, \lambda_{\mu}, \lambda_{\tau}).
\eeq
 Moreover, we assume that the RHN mass matrix $M_{N}$ is strictly degenerate at the GUT scale, which will be taken to be
 $M_G\simeq 2\cdot 10^{16}$~GeV.\footnote{Degeneracy of $M_N$ can be guaranteed by some symmetry at high energies. For concreteness, we
 assume
 this energy interval  to be $\geq M_G$ (although the degeneracy at lower energies can be considered as well).}
Therefore, we assume:
 \beq
{\rm at}~~\mu=M_G:~~~M_{N}= \left(\begin{array}{ccc}
 0&1\\
 1&0
\end{array}\right)M(M_G). \la{m01}
\eeq
 This form of $M_{N}$ is crucial for our studies.
  Although it is interesting and worth to study, we do not attempt here to justify the form of $M_N$ (and of the textures
 considered below) by symmetries. Our approach here is rather phenomenological aiming to investigate possibilities,
 outcomes and implications of the textures we consider.
 Since (\ref{m01}) at a tree level leads to the mass degeneracy of the RHN's, it has
interesting implications for resonant leptogenesis
\cite{{Branco:2006hz}, {Babu:2007zm}, {Babu:2008kp}} and also, as we will see
below, for building predictive neutrino scenarios \cite{Babu:2008kp}, \cite{Achelashvili:2016nkr}.

For the leptogenesis scenario two necessary conditions need to be satisfied. First of all, at the scale $\mu =M_{N_{1,2}}$ the degeneracy
between
the masses of $N_1$ and $N_2$ has to be lifted. And, at the same scale, the neutrino Yukawa matrix $\hat Y_{\nu }$ - written in the mass
eigenstate  basis of $M_N$, must be such that ${\rm Im}[ (\hat Y_{\nu }^\dag \hat Y_{\nu })_{12}]^2\neq 0$.  [These  can be seen
from Eq. (\ref{res-lept-asym}) with a demand $\ep_{1,2}\neq 0$.]
Below we show that both of them are realized by radiative corrections and needed effect already arises at 1-loop level, with
a dominant contribution due to the $Y_e$ Yukawa couplings (in particular from $\lam_{\tau }$ and in some cases from $\lambda_{\mu }$) in the
RG.

As it was shown \cite{early-tau-effect}, \cite{Babu:2008kp}, \cite{Achelashvili:2016trx}, within considered setup, radiative corrections are crucial for generating
cosmological CP violation.
In particular, the needed asymmetry is generated at 1-loop level due to $\lambda_{\tau}$ Yukawa coupling provided
that the condition
$(Y_{\nu })_{31}(Y_{\nu })_{32}\neq 0$ is satisfied \cite{Achelashvili:2016trx}. Here, to be more generic and to not limit the class of the models,  we also include the
effects of the $\lambda_{\mu}$ Yukawa coupling in the calculation.\footnote{In Sections \ref{Resonant} and \ref{improvement}, among other neutrino scenarios, we consider ones for
which such corrections are crucial for generation of the needed amount of Baryon asymmetry. } Thus, in this section we present details of
these calculations. We will start with radiative corrections to the $M_N$ matrix. RG effects cause lifting of the mass degeneracy and, as we
will see,
are important also for the phase misalignment (explained below).

At the GUT scale, the $M_N$ has off-diagonal form with $(M_N)_{11}=(M_N)_{22}=0$ [see Eq. (\ref{m01})]. However, at low energies, RG
corrections generate these entries. Thus, we parameterize the matrix $M_N$ at scale $\mu $ as:
\beq
M_{N}(\mu )=\left(
  \begin{array}{cc}
    \de_{N}^{(1)}(\mu ) & 1 \\
    1 & \de_{N}^{(2)}(\mu ) \\
  \end{array}
\right) \!M(\mu ).
\la{MN-with-deltas}
\eeq
While all entries of the matrix $M_N$ run, for our studies will be relevant the ratios $\fr{(M_N)_{11}}{(M_N)_{12}}=\de_{N}^{(1)}$
and $\fr{(M_N)_{22}}{(M_N)_{12}}=\de_{N}^{(2)}$ (obeying the RG equations investigated below). That's why $M_N$ was parametrized in a
form given
in Eq. (\ref{MN-with-deltas}).
With $|\de_{N}^{(1,2)}|\ll 1$, the $M$ (at scale $\mu =M$) will determine the masses of RHNs $M_1$ and $M_2$, while
$\de_{N}^{(1,2)}$ will be responsible for their splitting and for complexity in $M_N$ (the phase of the overall factor $M$ does not
contribute to the physical CP). As will be shown below:
\beq
 \de_{N}^{(1)}=(\de_{N}^{(2)})^*\equiv -\de_N .
 \la{rel-deltN}
 \eeq
 Therefore, $M_N$ is diagonalized by the transformation
$$
U_N^TM_NU_N=M_N^{Diag}={\rm Diag}\l M_1, M_2\r ~,~~~~~{\rm with}~~U_N=P_NO_N{P_N}' ~,
$$
\beq
M_1=|M|\l 1-|\de_N|\r ~,~~~~~~M_2=|M|\l 1+|\de_N|\r ~,
\la{MN-diag-tion}
\eeq
where
$$
P_N={\rm Diag}\l e^{-i\eta/2}, e^{i\eta/2}\r ,~~~
O_N=\fr{1}{\sq{2}}\left(
      \begin{array}{cc}
        1 & -1 \\
        1 & 1 \\
      \end{array}
    \right) ,
~~~{P_N}'={\rm Diag}\l e^{-i\phi_M/2}, ie^{-i\phi_M/2}\r ,
$$
\beq
{\rm with}~~~\eta ={\rm Arg}\l \de_N\r ~,~~~~~\phi_M={\rm Arg}\l M\r .
\la{P-ON-P1}
\eeq

In the $N$'s mass eigenstate basis, the Dirac type neutrino Yukawa  matrix will be $\hat Y_{\nu }=Y_{\nu }U_N$. In the CP asymmetries, the
components
$(\hat Y_{\nu }^\dag \hat Y_{\nu })_{21}$ and $(\hat Y_{\nu }^\dag \hat Y_{\nu })_{12}$ appear [see Eq. (\ref{res-lept-asym})].
From (\ref{MN-diag-tion}) and (\ref{P-ON-P1}) we have
\beq
\left [(\hat Y_{\nu }^\dag \hat Y_{\nu })_{21} \right ]^2=- \left [ (O_N^TP_N^*Y_{\nu }^\dag Y_{\nu }P_NO_N)_{21}\right ]^2 ,~~~~
\left [(\hat Y_{\nu }^\dag \hat Y_{\nu })_{12} \right ]^2=- \left [ (O_N^TP_N^*Y_{\nu }^\dag Y_{\nu }P_NO_N)_{12}\right ]^2 .
\la{YdagY-sq}
\eeq
Therefore, the CP violation should come from $P_N^*Y_{\nu }^\dag Y_{\nu }P_N$, which in a matrix form is:
\beq
P_N^*Y_{\nu }^\dag Y_{\nu }P_N=
\left(
  \begin{array}{cc}
    (Y_{\nu }^\dag Y_{\nu })_{11} & \left |(Y_{\nu }^\dag Y_{\nu })_{12}\right |e^{i(\eta -\eta')}\\
    |(Y_{\nu }^\dag Y_{\nu })_{21}|e^{i(\eta'-\eta)} & (Y_{\nu }^\dag Y_{\nu })_{22} \\
  \end{array}
\right) ,~~~~{\rm with}~~~~\eta'={\rm Arg} [(Y_{\nu }^\dag Y_{\nu })_{21}] ~.
\la{PYYP-form}
\eeq
We see that  $\eta'-\eta$ difference (mismatch) will govern the CP asymmetric decays of the RHNs. Without including the charged lepton Yukawa
couplings in the
RG effects we will have $\eta' \simeq \eta $ with a high accuracy. It was shown in Ref.  \cite{Dev:2015wpa} that by ignoring $Y_e$ Yukawas no
CP asymmetry emerges at ${\cal O}(Y_{\nu}^4)$ order and  non-zero contributions start only from ${\cal O}(Y_{\nu}^6)$ terms
\cite{Pilaftsis:2015bja}.
Such corrections are extremely suppressed for $Y_{\nu } \stackrel{<}{_\sim }1/50$.
Since in our consideration we are interested in cases with $M_{1,2}\stackrel{<}{_\sim }10^7$~GeV leading to $|(Y_{\nu })_{ij}|<7\cdot 10^{-4}$
(well fixed from the neutrino sector and the desired value of the baryon asymmetry), these effects
(i.e. order $\sim Y_{\nu}^6$ corrections) will not have any relevance.
In Ref.  \cite{Babu:2008kp}  in the RG of $M_N$  the effect
of $Y_e$, coming from 2-loop corrections,  was taken into account
and it was shown that sufficient CP violation can emerge. Below we show that including $Y_e$  in the  $Y_{\nu}$'s 1-loop RG,
will induce sufficient amount of CP violation. This mainly happens via $\lam_{\tau}$ and in particular cases (which are considered
below) from $\lambda_{\mu }$ Yukawa couplings. Thus, below we give detailed
investigation of
$\lambda_{\tau, \mu}$'s effects.

Using $M_N$'s RG given in Eq. (\ref{MN-2loop-RG}) (of Appendix \ref{app-YM-RGs}), for $\de_N^{(1,2)}$,
which are the ratios $\fr{(M_N)_{11}}{(M_N)_{12}}$
and $\fr{(M_N)_{22}}{(M_N)_{12}}$, [see parametrization in Eq. (\ref{MN-with-deltas})],
we can derive the following RG equations:
$$
16\pi^2 \fr{d}{dt}\de_N^{(1)}\!=\!4(Y_{\nu}^\dag Y_{\nu })_{21}\!+\!2\de_N^{(1)}\! \left [(Y_{\nu}^\dag Y_{\nu })_{11}\!-\!(Y_{\nu}^\dag
Y_{\nu })_{22}\right ] \!-\!
2(\de_N^{(1)})^2(Y_{\nu}^\dag Y_{\nu })_{12}\! -\! 2\de_N^{(1)}\de_N^{(2)} (Y_{\nu}^\dag Y_{\nu })_{21}
$$
\beq
-\fr{1}{4\pi^2}(Y_{\nu}^\dag Y_eY_e^\dag Y_{\nu })_{21}+\cdots
\la{deN1-RG}
\eeq
$$
16\pi^2 \fr{d}{dt}\de_N^{(2)}\!=\!4(Y_{\nu}^\dag Y_{\nu })_{12}\!+\!2\de_N^{(2)}\! \left [(Y_{\nu}^\dag Y_{\nu })_{22}\!-\!(Y_{\nu}^\dag
Y_{\nu })_{11}\right ] \!-\!
2(\de_N^{(2)})^2(Y_{\nu}^\dag Y_{\nu })_{21}\! -\! 2\de_N^{(1)}\de_N^{(2)} (Y_{\nu}^\dag Y_{\nu })_{12}
$$
\beq
-\fr{1}{4\pi^2}(Y_{\nu}^\dag Y_eY_e^\dag Y_{\nu })_{12}+\cdots
\la{deN2-RG}
\eeq
were in second lines of (\ref{deN1-RG}) and (\ref{deN2-RG}) are given 2-loop corrections depending on $Y_e$. Dots there stand for higher
order irrelevant terms. From 2-loop corrections we keep only $Y_e$ dependent terms. Remaining contributions
are not relevant for us.\footnote{Omitted terms  are either strongly suppressed or do not give any significant contribution to either the CP
violation or the RHN mass splittings.}
From  (\ref{deN1-RG}) and (\ref{deN2-RG}) we see that dominant contributions come from the first terms of the r.h.s. and from those given in
the second rows. Other terms give
contributions of order ${\cal O}(Y_{\nu}^4)$ or higher and thus will be ignored. At this approximation we have
\beq
\de_N^{(1)}(t)\simeq \de_N^{(2)*}(t)\equiv -\de_N(t)\simeq
-\fr{1}{4\pi^2} \int_{t}^{t_G}\!\!\!dt ~ \l Y_{\nu }^\dag ({\bf 1}-\fr{1}{16\pi^2}Y_eY_e^\dag )Y_{\nu }\r_{\!21}~
\la{approx-deN12}
\eeq
where $t=\ln \mu $, $t_G=\ln M_G$ and we have used the boundary conditions at the GUT scale  $\de_N^{(1)}(t_G)=\de_N^{(2)}(t_G)=0$.
For evaluation of the integral in (\ref{approx-deN12}) we need to know the scale dependence of $Y_{\nu }$ and $Y_e$. This is
 found in Appendix \ref{app-YM-RGs} by solving the RG equations for  $Y_{\nu }$ and $Y_e$.
Using Eqs. (\ref{approx-Ynu-sol}) and  (\ref{RG-factors}),
the integral of the matrix appearing in (\ref{approx-deN12}) can be written as:
\beq
\int_{t_M}^{t_G} \!\!Y_{\nu }^\dag ({\bf 1}-\fr{1}{16\pi^2}Y_eY_e^\dag ) Y_{\nu } dt \simeq \bar{\ka }(M)Y_{\nu G}^\dag \left(
  \begin{array}{ccc}
    1 & 0 & 0 \\
    0 & \bar r_{\mu}(M) & 0 \\
    0 & 0 & \bar r_{\tau }(M) \\
  \end{array}
\right)Y_{\nu G}
\la{int-matrix}
\eeq
where
\beq
 \bar r_{\tau }(M)=\fr{\int_{t_M}^{t_G}\!\! \ka(t)r_{\tau}(t)(1-\fr{\lam_{\tau }^2}{16\pi^2})dt}{\int_{t_M}^{t_G}\! \ka(t)dt}~,~~ \bar r_{\mu
 }(M)=\fr{\int_{t_M}^{t_G}\!\! \ka(t)r_{\mu}(t)(1-\fr{\lam_{\mu }^2}{16\pi^2})dt}{\int_{t_M}^{t_G}\! \ka(t)dt}~,~
 \bar{\ka }(M)=\int_{t_M}^{t_G}\!\! \ka(t)dt~,
\la{bar-r-kapa-2loop}
\eeq
\beq
 r_{\tau }(\mu )=\eta^2_{\tau}(\mu)~,~~~~~ r_{\mu }(\mu )=\eta^2_{\mu}(\mu)~,~~~~~\ka(\mu)=\eta^6_t(\mu)\eta^2_{g\nu }(\mu)~
\la{r-kapa}
\eeq
and we have ignored $\lam_{e }$ Yukawa couplings. For the definition of $\eta $-factors see Eq. (\ref{RG-factors}).
The $Y_{\nu G}$ denotes corresponding Yukawa matrix at scale $\mu =M_G$.
On the other hand, we have:
\beq
\left. (Y_{\nu }^\dag Y_{\nu })\right |_{\mu=M}\simeq \ka(M)Y_{\nu G}^\dag \left(
  \begin{array}{ccc}
    1 & 0 & 0 \\
    0 & r_{\mu }(M) & 0 \\
    0 & 0 & r_{\tau }(M) \\
  \end{array}
\right)Y_{\nu G}~.
\la{YYnu-M}
\eeq
(Derivations are given in Appendix \ref{app-YM-RGs}.)

Comparing (\ref{int-matrix}) with (\ref{YYnu-M}) we see that difference in these matrix structures (besides overall flavor universal RG
factors) is
in the RG factors $r_{\tau, \mu }(M)$ and $\bar r_{\tau, \mu }(M)$. Without the $\lam_{\tau, \mu}$ Yukawa couplings  these factors are equal
and there is no
mismatch between
the phases $\eta $ and $\eta'$ [defined in Eqs. (\ref{P-ON-P1}) and  (\ref{PYYP-form})] of these matrices. Non zero $\eta' -\eta $ will be
due to the deviations, which we parameterize as
\beq
\xi_{\tau} =\fr{\bar r_{\tau }(M)}{r_{\tau }(M)}-1,~~~~\xi_{\mu} =\fr{\bar r_{\mu }(M)}{r_{\mu }(M)}-1 ~.
\la{xi-shift}
\eeq
The values of
$\xi_{\mu }$ and $\xi_{\tau }$ can be computed numerically by evaluation of the appropriate RG factors. Approximate expressions can be
derived for $\xi_{\tau, \mu} $, which are given by:
$$
\xi_{\tau} \!\simeq  \!\left [\fr{\lam_{\tau}^2(M)}{16\pi^2}\ln \fr{M_G}{M}
+\fr{1}{3}\fr{\lam_{\tau}^2(M)}{(16\pi^2)^2}
\left [ 3\lam_t^2+6\lam_b^2+10\lam_{\tau}^2-(2c_e^a+c_{\nu}^a)g_a^2\right ]_{\mu=M} \l \!\ln \fr{M_G}{M} \!\r^{\!\!2}\right ]_{\rm 1-loop}
$$
\beq
-~\left [\! \fr{\lam^2_{\tau }(M)}{16\pi^2}\!\right ]_{\rm 2-loop}~,
\la{approx-for-xi}
\eeq
\\
$$
\xi_{\mu} \!\simeq  \!\left [\fr{\lam_{\mu}^2(M)}{16\pi^2}\ln \fr{M_G}{M}
+\fr{1}{3}\fr{\lam_{\mu}^2(M)}{(16\pi^2)^2}
\left [ 3\lam_t^2+6\lam_b^2+2\lambda^{2}_{\tau}-(2c_e^a+c_{\nu}^a)g_a^2\right ]_{\mu=M} \l \!\ln \fr{M_G}{M} \!\r^{\!\!2}\right ]_{\rm
1-loop}
$$
\beq
-~\left [\! \fr{\lam^2_{\mu }(M)}{16\pi^2}\!\right ]_{\rm 2-loop}~,
\la{approx-for-xi-mu}
\eeq
where one and two loop contributions are indicated. Derivation of approximate expression of $\xi_{\tau }$ [Eq.(\ref{approx-for-xi})] is given
in Appendix A.1 of Ref.\cite{Achelashvili:2016trx}. Eq. (\ref{approx-for-xi-mu}) can be derived in a similar way.
As we see, non-zero $\xi_{\tau, \mu} $ are induced already at 1-loop
level [without 2-loop correction of $\fr{\lam_{\tau, \mu }^2}{16\pi^2}$ in Eq. (\ref{bar-r-kapa-2loop})]. However, inclusion of 2-loop
correction
can contribute to the $\xi_{\tau, \mu} $ by amount of $\sim 3-5\%$ (because of $\ln \fr{M_G}{M}$ factor suppression) and we have included it.

Now we write down quantities which have direct relevance for leptogenesis calculations. Using Eq. (\ref{int-matrix}) in (\ref{approx-deN12})
and then applying Eq.(\ref{approx-Ynu-sol}) [for expressing $Y_{\nu G}$'s elements with corresponding entries of $Y_{\nu }(M)$], with
definitions of Eqs. (\ref{r-kapa}) and (\ref{xi-shift}), we obtain:
\beq
|\de_N(M) |e^{i\eta }=\fr{1}{4\pi^2}\fr{\bar \ka (M)}{\ka (M)}\left [ |(Y_{\nu}^\dag Y_{\nu})_{21}|e^{i\eta' } +\xi_{\tau}
|(Y_{\nu})_{31}(Y_{\nu})_{32}|e^{i(\phi_{31}-\phi_{32})}+\xi_{\mu}
|(Y_{\nu})_{21}(Y_{\nu})_{22}|e^{i(\phi_{21}-\phi_{22})}\right ]_{\mu=M}
\la{eta-vs-eta1}
\eeq
where $\phi_{ij}$ denotes the phase of the matrix element $(Y_{\nu })_{ij}$ at scale $\mu =M$. Eq.
(\ref{eta-vs-eta1}) shows well that
in the limit $\xi_{\tau, \mu} \to 0$, we have $\eta =\eta '$, while the mismatch between these two phases is due to $\xi_{\tau, \mu} \neq 0$.
With $\xi_{\tau, \mu} \ll 1$, from
(\ref{eta-vs-eta1})
we derive:
\beq
\eta -\eta' \simeq \fr{\xi_{\tau}|(Y_{\nu})_{31}(Y_{\nu})_{32}|\sin (\phi_{31}-\phi_{32}-\eta')+\xi_{\mu}|(Y_{\nu})_{21}(Y_{\nu})_{22}|\sin
(\phi_{21}-\phi_{22}-\eta')}{|(Y_{\nu}^\dag Y_{\nu})_{21}|} ~.
\la{eta-eta1-aprox}
\eeq
We stress, that the 1-loop renormalization of the $Y_{\nu }$ matrix plays the leading role in generation of $\xi_{\tau, \mu}$, i.e. in the
CP violation.\footnote{Note that since RG equations for $M_N$ and $Y_{\nu}$ in non-SUSY case have similar structures
(besides some group-theoretical factors) the $\xi_{\tau, \mu} $ would be generated also within non-SUSY setup.}
[This is also demonstrated by Eq. (\ref{approx-for-xi}).] When the product $(Y_{\nu})_{31}(Y_{\nu})_{32}$ is non-zero, the leading role for
the mismatch between $\eta $ and $\eta'$ is played by $\xi_{\tau}$. However, for the Yukawa texture, having this  product zero, important
will be contribution from $\xi_{\mu }$. [As we will see on working examples, this will happen for $T_9$ of Eq. (\ref{xxx}) and texture
$B_2$ of Eq. (\ref{textureB12})].

The value of $|\de_N(M) |$, which characterizes the mass splitting between the RHN's, can be computed by taking the absolute values of both sides of
(\ref{eta-vs-eta1}):
\beq
|\de_N(M) |=\fr{\ka_N}{4\pi^2}\left |(Y_{\nu}^\dag Y_{\nu})_{21} +\xi_{\tau} (Y_{\nu})_{31}(Y_{\nu}^*)_{32}+\xi_{\mu}
(Y_{\nu})_{21}(Y_{\nu}^*)_{22} \right |_{\mu=M}\ln
\fr{M_G}{M}~,~{\rm with}~~
\ka_N=\fr{\bar{\ka}(M)}{\ka (M)\ln \fr{M_G}{M}}~.
\la{abs-deltaN}
\eeq
These expressions can be used upon the calculation of the leptogenesis, which we will do in sections \ref{Resonant} and \ref{improvement} for concrete
models of the neutrino mass matrices.
\section{See-Saw via Two Texture Zero $3\times 2$ Dirac Yukawas Augmented by Single d=5 Operator. Predicting CP Violation}
\la{RG-CP-1}
Within the setup with two RHNs, having at the GUT scale mass matrix of the form  (\ref{m01}), we consider all two texture zero $3\times 2$
Yukawa matrices. As given in \cite{Achelashvili:2016nkr}, there are nine such different matrices:
\beqs
T_{1}= \left(\begin{array}{ccc}
\times&0\\
\times&0\\
\times&\times
\end{array}\right),\quad
T_{2}= \left(\begin{array}{ccc}
\times&0\\
\times&\times\\
\times&0
\end{array}\right),\quad
T_{3}= \left(\begin{array}{ccc}
\times&\times\\
\times&0\\
\times&0
\end{array}\right),
\eeqs
\beqs
T_{4}= \left(\begin{array}{ccc}
0&0\\
\times&\times\\
\times&\times
\end{array}\right),\quad
T_{5}= \left(\begin{array}{ccc}
\times&0\\
0&\times\\
\times&\times
\end{array}\right),\quad
T_{6}= \left(\begin{array}{ccc}
\times&0\\
\times&\times\\
0&\times
\end{array}\right),
\eeqs
\beq
T_{7}= \left(\begin{array}{ccc}
\times&\times\\
0&0\\
\times&\times
\end{array}\right),\quad
T_{8}= \left(\begin{array}{ccc}
\times&\times\\
\times&0\\
0&\times
\end{array}\right),\quad
T_{9}= \left(\begin{array}{ccc}
\times&\times\\
\times&\times\\
0&0
\end{array}\right),\la{xxx}
\eeq
 where "$\times$"s stand for non-zero entries. From these textures one can factor out phases in such a way as to make maximal number of
 entries be real. As it was shown in \cite{Achelashvili:2016nkr}, phases can be removed from all textures  besides $T_4, T_7$ and $T_9$. Thus,
 here we pick up only $T_{4, 7, 9}$ textures, which lead to cosmological CP violation and have potential to realize resonant leptogenesis
 \cite{Babu:2007zm}, \cite{Babu:2008kp} (due to quasi-degenerate $N_1$ and $N_2$ states). Therefore, we can parametrize these three textures
 as:
\\
\\
TEXTURE $T_{4}$
\beq
T_{4}=\begin{pmatrix}
0 & 0\\
a_{2}e^{i\alpha_{2}} & b_{2}e^{i\beta_{2}}\\
a_{3}e^{i\alpha_{3}} &  b_{3}e^{i\beta_{3}}

\end{pmatrix}
=
\begin{pmatrix}
e^{ix} & 0&0\\
0 & e^{iy}&0\\
0 & 0&e^{iz}
\end{pmatrix}
\begin{pmatrix}
0 & 0\\
a_{2} & b_{2}\\
a_{3} &b_{3}e^{i\phi}
\end{pmatrix}
\begin{pmatrix}
e^{i\omega} & 0\\
0 & e^{i\rho}
\end{pmatrix}, \la{t4212}
\eeq
{\rm with}
\beq
\omega=\alpha_{2}-\beta_{2}+\rho,\quad y= \beta_{2}-\rho,\quad z= \alpha_{3}-\alpha_{2}+\beta_{2}-\rho, \quad
\phi=\alpha_{2}-\alpha_{3}+\beta_{3}-\beta_{2}. \la{t42121}
\eeq
\\
TEXTURE $T_{7}$
\beq
T_{7}=\begin{pmatrix}
a_{1}e^{i\alpha_{1}} & b_{1}e^{i\beta_{1}}\\
0 & 0\\
a_{3}e^{i\alpha_{3}} &  b_{3}e^{i\beta_{3}}
\end{pmatrix}
=
\begin{pmatrix}
e^{ix} & 0&0\\
0 & e^{iy}&0\\
0 & 0&e^{iz}
\end{pmatrix}
\begin{pmatrix}
a_{1} & b_{1}\\
0 & 0\\
a_{3} &b_{3}e^{i\phi}
\end{pmatrix}
\begin{pmatrix}
e^{i\omega} & 0\\
0 & e^{i\rho}
\end{pmatrix}, \la{t7218}
\eeq
{\rm with}
\beq
\omega= \rho+\alpha_{1}-\beta_{1}, \quad x= \beta_{1}-\rho, \quad
z= \alpha_{3}-\alpha_{1}+\beta_{1}-\rho,\quad \phi=
\alpha_{1}-\alpha_{3}-\beta_{1}+\beta_{3}. \la{t72181}
\eeq
\\
TEXTURE $T_{9}$
\beq
T_{9}=\begin{pmatrix}
a_{1}e^{i\alpha_{1}} & b_{1}e^{i\beta_{1}}\\
a_{2}e^{i\alpha_{2}} & b_{2}e^{i\beta_{2}}\\
0 &  0
\end{pmatrix}
=
\begin{pmatrix}
e^{ix} & 0&0\\
0 & e^{iy}&0\\
0 & 0&e^{iz}
\end{pmatrix}
\begin{pmatrix}
a_{1} & b_{1}\\
a_{2} &  b_{2}e^{i\phi}\\
0 &0
\end{pmatrix}
\begin{pmatrix}
e^{i\omega} & 0\\
0 & e^{i\rho}
\end{pmatrix}, \la{t9222}
\eeq
{\rm with}
\beq
\omega=\alpha_{1}-\beta_{1}+\rho, \quad x= \beta_{1}-\rho, \quad
y= \alpha_{2}-\alpha_{1}+\beta_{1}-\rho, \quad \phi=
\alpha_{1}-\beta_{1}-\alpha_{2}+\beta_{2}. \la{t9223}
\eeq
The phases $x, y$ and $z$ can be eliminated  by proper redefinition of
the states $l$ and $e^c$.
 As far as the phases  $\omega $ and $\rho $ are concerned,
because of the form of the
$M_N$ matrix (\ref{m01}), they too
 will turn out to be non-physical.
As we see, in textures $T_{4}$, $T_{7}$ and $T_{9}$ there remains
one unremovable phase $\phi$  (i.e. in the second matrices of the r.h.s. of Eqs. (\ref{t4212})
(\ref{t7218}) and (\ref{t9222}) respectively).  This physical phase $\phi$ is relevant to the leptogenesis\cite{Babu:2008kp} and also, as it was
shown in \cite{Achelashvili:2016nkr}, it can be related to phase $\delta$, determined from the neutrino sector.
As will be shown on  concrete neutrino models, this will remain true after inclusion of specific single $d=5$ operator.
Integrating the RHN's, from the superpotential couplings of Eq. (\ref{r21}), using the see-saw formula, we get the following contribution to
the light neutrino mass matrix:
\beq
  M^{ss}_{\nu}=-\langle h^{0}_{u}\rangle^{2} Y_{\nu}M^{-1}_{N}Y^{T}_{\nu}. \la{seesaw}
\eeq
For $Y_{\nu }$ in (\ref{seesaw}) the textures $T_{4, 7, 9}$
 should be used in turn. All obtained matrices
$M_{\nu }^{ss}$, if identified with light neutrino mass matrices, will give experimentally unacceptable results. The reason is the number of
texture zeros which we have in $T_{i}$ and $M_{N}$ matrices. In order to overcome this difficulty, in Ref. \cite{Achelashvili:2016nkr}, the
following single $d=5$ operator was included for each case:
 \beq
{\cal
O}^{5}_{ij}\equiv\frac{\tilde{d_{5}}e^{i{x_{5}}}}{2M_{*}}l_{i}l_{j}h_{u}h_{u}
\la{d5}
\eeq where $\tilde{d_{5}}$, $x_{5}$ and $M_{*}$ are real parameters. (\ref{d5}), together with (\ref{seesaw}) will contribute to the neutrino
mass matrix. This will allow to have viable models and,
at the same time because of the minimal number of the additions, we will still have predictive scenarios. The operators (\ref{d5}) can be
obtained by another sector in such a way as to not affect the forms of $T_{4, 7, 9}$ and $M_{N}$ matrices (one detailed example was presented
in \cite{Achelashvili:2016trx}). See Sect. \ref{discussions}
 for more discussion on a
 possible origin of the (\ref{d5}) type operators. Above we have written the Yukawa textures in the form: \beq
Y_{\nu}={\cal P}_1Y^{R}_{\nu}{\cal P}_2, \eeq where ${\cal P}_1, {\cal P}_2$ are diagonal phase
matrices  and $Y^{R}_{\nu}$  contains
only one phase. Making the field
phase redefinitions:
\beq
l^{\prime}={\cal P}_1l, \quad N^{\prime}={\cal P}_2N, \quad (e^{\prime})^{c}={\cal P}^{*}_{1}e^{c}~~ \mathrm{with}  ~~{\cal
P}_1=\mathrm{Diag} (e^{ix}, e^{iy}, e^{iz}),~~~{\cal P}_2=\mathrm{Diag} (e^{i \omega }, e^{i \rho })
\eeq
the superpotential coupling will become:
 \begin{equation}
W_{e}=(l^{\prime})^{T}Y_{e}^{\rm diag}(e^{\prime})^{c}h_{d},\quad
W_{\nu}=(l^{\prime})^{T}Y^{R}_{\nu}N^{\prime}h_{u}-\frac{1}{2}(N^{\prime})^{T}M^{\prime}_{N}N^{\prime}
\end{equation}
with:
\beq M^{\prime}_{N}=
\begin{pmatrix}
0 & 1\\
1 & 0
\end{pmatrix}M e^{-i(\omega + \rho)}  \la{M-prime-N}.
\eeq
Now, for simplification of the notations, we will get rid of the primes (i.e. perform $l^{\prime}\rightarrow l$, $e^{c \prime}\rightarrow
e^{c}$,...) and in Eq. (\ref{seesaw}) using $Y_{\nu}^R$ instead of $Y_{\nu}$, from different $T_{4, 7, 9}$ textures we get corresponding
$M_{\nu}^{ss}$, and then adding the single operator (\ref{d5}) terms to zero entries of (\ref{seesaw}), one per $M_{\nu }^{ss}$, obtain the final neutrino mass matrices.
Doing so, one obtains the neutrino mass matrices \cite{Achelashvili:2016nkr}:
\beq
P_{1}=\left(\begin{array}{ccc}
0 & \times&0\\
\times& \times&\times\\
0&\times&\times
\end{array}\right),\quad
P_{2}=\left(\begin{array}{ccc}
0& 0&\times\\
0&\times&\times\\
\times&\times&\times
\end{array}\right),\quad
P_{3}=\left(\begin{array}{ccc}
\times &0&\times\\
0& 0&\times\\
\times&\times&\times
\end{array}\right),\quad
P_{4}=\left(\begin{array}{ccc}
\times& \times&0\\
\times& \times&\times\\
0&\times&0
\end{array}\right), \la{pse}
\eeq
where each type of texture originate as:
$$
P_{1}-{\rm type}:\quad M^{(12)}_{T_{4}},\qquad P_{2}-{\rm type}:\quad M^{(13)}_{T_{4}},\quad P_{3}-{\rm type}:\quad M^{(23)}_{T_{7}} ,\qquad P_{4} -{\rm type}:\quad M^{(23)}_{T_{9}}
$$
where subscript for $M$ indicates which Yukawa texture the see-saw part [of Eq. (\ref{seesaw})] came from, while superscript denotes the non-zero
mass matrix element arising from the addition of the $d=$5 operator of type (\ref{d5}).
Since within our setup we are deriving neutrino mass matrices, we are able to renormalize them from high scales down to $M_Z$. With details
given in the Appendix A of Ref. \cite{Achelashvili:2016trx}, we here write down $P_{1,2,3,4}$ textures at scale $M_Z$ and give results
already obtained in \cite{Achelashvili:2016nkr}. Before doing this, we set up conventions, which are used below. Since we work in the basis
in which charged lepton Yukawa matrix is diagonal and real, the lepton mixing matrix $U$ is related to the neutrino mass matrix as:
\beq
M_{\nu}=PU^{*}P^{'}M_{\nu}^{\rm diag}U^{+}P \la{nu1}
\eeq
where  $M_{\nu}^{\rm diag}=(m_{1},m_{2},m_{3})$ ($m_{1,2,3}$ are light neutrino masses)
and the phase matrices and $U$ are:
\beq
 P={\rm
Diag}(e^{i\omega_{1}},e^{i\omega_{2}},e^{i\omega_{3}}),\quad
P^{'}={\rm Diag}(1,e^{i\rho_{1}},e^{i\rho_{2}}),\la{nu2}
\eeq \beq
U= \left(\begin{array}{ccc}
c_{13}c_{12} &c_{13}s_{12}&s_{13}e^{-i\delta}\\
-c_{23}s_{12}-s_{23}s_{13}c_{12}e^{i\delta}&
c_{23}c_{12}-s_{23}s_{13}s_{12}e^{i\delta}&s_{23}c_{13}\\
s_{23}s_{12}-c_{23}s_{13}c_{12}e^{i\delta}&-s_{23}c_{12}-c_{23}s_{13}s_{12}e^{i\delta}&c_{23}c_{13}
\end{array}\right)\la{nu3},
\eeq
\\
where $s_{ij}\equiv \sin \theta_{ij}$ and $c_{ij}\equiv \cos \theta_{ij}$. For normal and inverted neutrino mass orderings (denoted
respectively by NH and IH) we will use notations:
\beq
\Delta
m_{sol}^{2}=m_{2}^{2}-m_{1}^{2},\quad
\Delta
m_{atm}^{2}=m_{3}^{2}-m_{2}^{2},\quad
m_{1}=\sqrt{m_{3}^{2}-\Delta m_{atm}^{2}-\Delta m_{sol}^{2}},\quad
m_{2}=\sqrt{m_{3}^{2}-\Delta m_{atm}^{2}} \la{nh1}
\eeq
\\
\beq
\Delta
m_{atm}^{2}=m_{2}^{2}-m_{3}^{2},\quad
\Delta
m_{sol}^{2}=m_{2}^{2}-m_{1}^{2},\quad
m_{1}=\sqrt{m_{3}^{2}+\Delta m_{atm}^{2}-\Delta m_{sol}^{2}},\quad
m_{2}=\sqrt{m_{3}^{2}+\Delta m_{atm}^{2}} \la{ih1}
\eeq
As far as the numerical values of the oscillation parameters are concerned, since the bfv's of the works of Ref. \cite{recent-nu-data} differ from each other by few \%'s, we will use their mean values:
$$
\sin^2\theta_{12}=0.308 ,~~~~~~
\sin^2\theta_{23}=\left\{
                                                   \begin{array}{ll}
                                                     \!\!0.432 &\! \hbox{for NH} \\
                                                     \!\!0.591 &\! \hbox{for IH}
                                                   \end{array}
                                                 \right. ,~~~
\sin^2\theta_{13}=\left\{
                                                   \begin{array}{ll}
                                                     \!\! 0.02157 &\! \hbox{for NH} \\
                                                     \!\! 0.0216 &\! \hbox{for IH}
                                                   \end{array}
                                                 \right. ,
$$
\beq
\Delta m^2_{sol}=7.48\cdot 10^{-5}~{\rm eV}^2 ,~~~~
\Delta m^2_{atm}=|m_3^2-m_2^2|=\left\{
                                                   \begin{array}{ll}
                                                     \!\! 2.47\cdot 10^{-3}~{\rm eV}^2 &\! \hbox{for NH} \\
                                                     \!\! 2.54\cdot 10^{-3}~{\rm eV}^2 &\! \hbox{for IH}
                                                   \end{array}
                                                 \right. .
\la{ever-bfv}
\eeq
In models, which allow to do so, we use the best fit values (bfv) given in (\ref{ever-bfv}). However, in some cases
we also apply the value(s) of some oscillation parameter(s) which deviate from the bfv's by several $\sigma $.
\begin{center}
{\centering \textbf{$P_1$ Neutrino Texture}}
\end{center}
This texture, within our scenario, can be parameterized as:
\beq
M_{\nu}(M_Z)=\begin{pmatrix}
0&d_{5}&0\\
d_{5}&2a_2b_2&(a_3b_2+a_2b_3e^{i\phi})r_{\nu 3}\\
0&(a_3b_2+a_2b_3e^{i\phi})r_{\nu 3}&2a_3b_3e^{i\phi}r_{\nu 3}^2
\end{pmatrix}\bar{m} \la{M-p1}
\eeq
where,
\beq
 \bar m=-\frac{r_{\bar{m}}v_{u}^2(M_Z)}{M\cdot e^{-i(\omega+\rho)}} \la{bar-m-rbar}
\eeq
\begin{center}
 \begin{tabular}{|l|r|r|r|}
  \hline
  \multicolumn{1}{|c|}{\sffamily $\delta$}
 &\multicolumn{1}{|c|}{\sffamily $\rho_{1}$}&\multicolumn{1}{|c|}{\sffamily $\rho_{2}$}&\multicolumn{1}{|c|}{\sffamily works with}\\
  \hline
  $\pm 0.0879121$&$\pm 3.11851$&$\pm 3.03949$&\makecell{NH, $\sin^{2}\theta_{23}=0.451$, $\sin^{2}\theta_{12}=0.323$ and best\\ fit values for remaining
  oscillation parameters,\\ $(m_{1},m_{2},m_{3})=(0.00694406,0.0110914,0.0509217)$, $m_{\beta\beta}=0$}\\
  \hline
  \end{tabular}
  \captionof{table}{Results from $P_{1}$ type texture. Masses are given in eVs.}
  \label{tab01}
  \end{center}
and RG factors $r_{\bar m}$ and $r_{\nu 3}$ are given in Eqs. (A.17) and (A.18) of Ref. \cite{Achelashvili:2016trx}.
(For notations and definitions see also Appendix \ref{app-nuRG} of the present paper.)
The entries depending on $a_i$, $b_j$ in (\ref{M-p1}) arise from the $T_4$ texture [given in (\ref{t4212})] by the see-saw mechanism.
 The entry $d_5$ comes from the (\ref{d5}) type operator
 $\frac{\tilde d_5e^{ix_5}}{M_*}l_1l_2h_uh_u$. Since, as we see from Eqs. (\ref{t4212}) and (\ref{t42121}), the phase $x$ is undetermined, we can select it in such a way as to set (\ref{M-p1})'s $d_5$ entry to be real. Therefore, we still have single physical phase $\phi $. It will be related to the phase
 $\delta $ and will govern the leptogenesis process (discussed in Sect. \ref{Resonant}).
Due to the texture zeros, it is possible to predict the phases and values of the neutrino masses in terms of the measured oscillation
parameters.
In particular, the conditions $M^{(1,1)}_{\nu}=0$ and $M^{(1,3)}_{\nu}$=0, using (\ref{nu1})-(\ref{nu3}), give:
\beq
\frac{m_{1}}{m_{3}}c^{2}_{12}+\frac{m_{2}}{m_{3}}s^{2}_{12}e^{i\rho_{1}}=-t^{2}_{13}e^{i(\rho_{2}+2\delta)}
\la{rels-fromP1-1}
\eeq
and
\beq
-\left(\frac{m_{1}}{m_{3}}-\frac{m_{2}}{m_{3}}e^{i\rho_{1}}\right)t_{23}s_{12}c_{12}-s_{13}e^{i(\rho_{2}+
\delta)}+s_{13}e^{-i\delta}\left(\frac{m_{1}}{m_{3}}c^{2}_{12}+\frac{m_{2}}{m_{3}}s^{2}_{12}e^{i\rho_{1}}\right)=0 .
\la{rels-fromP1-2}
\eeq
These two complex equations with the input of five oscillation parameters allow to calculate all neutrino masses and predict three phases $\de , \rho_1$ and $\rho_2$. Without providing here further analytical relations
[followed from Eqs. (\ref{rels-fromP1-1}), (\ref{rels-fromP1-2}) and  given in \cite{Achelashvili:2016nkr}), in Table \ref{tab01} we summarize the results. [Only normal hierarchical (NH)
neutrino mass ordering scenario works for the $P_1$ type texture.]
\begin{center}
{\centering \textbf{$P_2$ Neutrino Texture}}
\end{center}
\beq
M_{\nu}(M_Z)=\begin{pmatrix}
0&0&d_{5}\\
0&2a_2b_2&(a_3b_2+a_2b_3e^{i\phi})r_{\nu 3}\\
d_{5}&(a_3b_2+a_2b_3e^{i\phi})r_{\nu 3}&2a_3b_3e^{i\phi}r_{\nu 3}^2
\end{pmatrix}\bar{m} \la{M-p2}
\eeq
This texture's $a_i, b_i$ entries are also obtained from the $T_4$ texture (\ref{t4212}) via the see-saw mechanism and
by addition of the $d=5$ operator $\frac{\tilde d_5e^{ix_5}}{M_*}l_1l_3h_uh_u$.
By proper adjustment of the phase $x$ [remaining undetermined in (\ref{t4212}) and (\ref{t42121})], we  can set $d_5$ entry of (\ref{M-p2}) to be real.
The two conditions $M^{(1,1)}_{\nu}=0$ and $M^{(1,2)}_{\nu}$=0 give relation of Eq. (\ref{rels-fromP1-1}) and
\beq
-\left(\frac{m_{1}}{m_{3}}-\frac{m_{2}}{m_{3}}e^{i\rho_{1}}\right)s_{12}c_{12}+s_{13}t_{23}e^{i(\rho_{2}+
\delta)}-s_{13}t_{23}e^{-i\delta}\left(\frac{m_{1}}{m_{3}}c^{2}_{12}+\frac{m_{2}}{m_{3}}s^{2}_{12}e^{i\rho_{1}}\right)=0.
\la{rels-fromP2-2}
\eeq
which allow to predict neutrino masses and three phases $\de, \rho_{1,2}$.
Results are given in Table \ref{tab02}. For inputs the best fit values (bfv) of the oscillation parameters are taken from
Eq.(\ref{ever-bfv}). For more details we refer the reader to \cite{Achelashvili:2016nkr}.
 \begin{center}
  \begin{tabular}{|l|r|r|r|}
  \hline
  \multicolumn{1}{|c|}{\sffamily $\delta$}
 &\multicolumn{1}{|c|}{\sffamily $\rho_{1}$}&\multicolumn{1}{|c|}{\sffamily $\rho_{2}$}&\multicolumn{1}{|c|}{\sffamily works with}\\
  \hline
  $\pm 1.71006$&$\mp  2.79206$&$\mp 1.47308$&\makecell{NH and bfv's\\ of oscillation parameters,\\
  $(m_{1},m_{2},m_{3})=(0.00471158,0.0098488,0.0506656)$, $m_{\beta\beta}=0$}\\
  \hline
  \end{tabular}
  \captionof{table}{Results from $P_{2}$ type texture. Masses are given in eVs.}
  \label{tab02}
  \end{center}
\begin{center}
{\centering \textbf{$P_3$ Neutrino Texture}}
\end{center}
Using the see-saw formula (\ref{seesaw}) for the $T_7$ texture (\ref{t7218}) and including the $d=5$ operator
$\frac{\tilde d_5e^{ix_5}}{M_*}l_2l_3h_uh_u$, we obtain the $P_3$ neutrino texture:
\beq
M_{\nu}(M_Z)=\begin{pmatrix}
2a_1b_1&0&(a_3b_1+a_1b_3e^{i\phi})r_{\nu 3}\\
0&0&d_5\\
(a_3b_1+a_1b_3e^{i\phi})r_{\nu 3}&d_5&2a_3b_3e^{i\phi}r_{\nu 3}^2
\end{pmatrix}\bar{m} \la{M-p3}
\eeq
Since the phase $y$ is not fixed in
 (\ref{t7218}) and (\ref{t72181}), without loss of any generality the $d_5$ entry of (\ref{M-p3}) can be set to be real.
The conditions  $M^{(1,2)}_{\nu}=0$ and $M^{(2,2)}_{\nu}$=0, similar to previous cases, allow to predict $m_{1,2,3}$ and $\de, \rho_{1,2}$.
Without giving the expressions (being lengthy and presented in Ref. \cite{Achelashvili:2016nkr}),
we proceed to give numerical results, which for
 NH and inverted hierarchical (IH) neutrino mass orderings are summarized in Table \ref{tab03}.
   \begin{center}
  \begin{tabular}{|l|r|r|r|}
  \hline
  \multicolumn{1}{|c|}{\sffamily $\delta$}
 &\multicolumn{1}{|c|}{\sffamily $\rho_{1}$}&\multicolumn{1}{|c|}{\sffamily $\rho_{2}$}&\multicolumn{1}{|c|}{\sffamily works with}\\
  \hline
 $\pm 1.53714$&$\pm 0.0867342$&$\pm 3.20236$&\makecell{NH and bfv's\\ of oscillation parameters,\\
 $(m_{1},m_{2},m_{3})=$\\$(0.0588907,0.0595224,0.077543)$,\\ $m_{\beta\beta}=0.059436$}\\
  \hline
  $\pm 1.58066$&$\mp 0.114316$&$\pm 3.06301$&\makecell{IH and bfv's\\ of oscillation parameters,\\
  $(m_{1},m_{2},m_{3})=$\\$(0.0696426,0.0701776,0.0488354)$,\\ $m_{\beta\beta}=0.0692588$}\\
  \hline
  \end{tabular}
  \captionof{table}{Results from $P_{3}$ type texture. Masses are given in eVs.}
  \label{tab03}
  \end{center}
\begin{center}
{\centering \textbf{$P_4$ Neutrino Texture}}
\end{center}
This texture is obtained by applying the see-saw formula (\ref{seesaw}) to the $T_9$ texture (\ref{t9222}) and including the $d=5$ operator
$\frac{\tilde d_5e^{ix_5}}{M_*}l_2l_3h_uh_u$. Doing these we obtain the $P_4$ neutrino texture:
\beq
M_{\nu}(M_Z)=\begin{pmatrix}
2a_1b_1&(a_2b_1+a_1b_2e^{i\phi})&0\\
(a_2b_1+a_1b_2e^{i\phi})&2a_2b_2e^{i\phi}&d_{5}\\
0&d_{5} &0
\end{pmatrix}\bar{m} \la{M-p4}
\eeq
In this case the phase $z$ is not fixed
[see Eqs. (\ref{t9222}) and (\ref{t9223})] and we can use this phase freedom to take  $d_5$ entry of (\ref{M-p4}) matrix as a real parameter.
The conditions $M_{\nu }^{(1,3)}=M_{\nu }^{(3,3)}=0$ will give two complex (i.e. four real) equations, which contain three phases
$\de, \rho_{1, 2}$ and one of the neutrino masses (remember that two measured parameters $\De m_{sol}^2=m_2^2-m_1^2$ and
$\De m_{atm}^2=|m_3^2-m_2^2|$ leave undetermined values of the neutrino masses).
Therefore, as for previous cases, with input of five measured oscillation parameters (which are: $\De m_{sol}^2, \De m_{atm}^2$ and
$\{\te_{12}, \te_{23}, \te_{13}\}$) from the conditions given above we predict all light neutrino masses and
three phases $\de, \rho_{1, 2}$.  Still referring to \cite{Achelashvili:2016nkr}, for analytical expressions,
in Table  \ref{tab04} we give the numerical results obtained for this texture $P_4$ for NH and IH cases. The value of $s^2_{23}$
we are using is deviated from the bfv, because the conditions
$M_{\nu }^{(1,3)}=M_{\nu }^{(3,3)}=0$ do not allow to use bfv's. Note
that in NH, case 2 and for IH the values
of $s^2_{23}$ are less deviated from bfv, but the NH's case 1, as it turns out, is preferred for obtaining needed amount of the baryon asymmetry. Without the latter constraint, just for satisfying the neutrino data, we could have used smaller values of $s^2_{23}$, but this would give higher  values of neutrino masses which would not satisfy the current cosmological constraint
$\sum_i m_i<0.23$~eV (the limit set by the Planck observations
\cite{Ade:2015xua}\footnote{Tighter upper bound can be obtained by considering additional combined datasets \cite{Vagnozzi:2017ovm}. However,
bound also depends on the theoretical framework and can be relaxed (see e.g. $2^{\rm nd}$ Ref. of \cite{recent-nu-data}, where as demonstrated in Table II, the scenario with extra $A_{\rm lens}$ parameter yields more relaxed bounds). Thus, upon our calculations we use the constraint $\sum_i m_i<0.23$~eV.}). Upon leptogenesis investigation we will use NH, case 1 given in Tab.\ref{tab04}.
\begin{center}
   \begin{tabular}{|c|l|r|r|r|}
  \hline
 \multicolumn{1}{|c|}{\sffamily $$}
 & \multicolumn{1}{|c|}{\sffamily $\delta$}
 &\multicolumn{1}{|c|}{\sffamily $\rho_{1}$}&\multicolumn{1}{|c|}{\sffamily $\rho_{2}$}&\multicolumn{1}{|c|}{\sffamily works with}\\
 \hline
NH, case 1&$\pm 1.62446$&$\mp 0.129186$&$\pm 3.05085$&\makecell{NH and $\sin^2 \theta_{23}=0.6$ and bfv's\\ for remaining
 oscillation parameters,\\ $(m_{1},m_{2},m_{3})=$\\$(0.044819,0.0456458,0.0674799)$,\\ $m_{\beta\beta}=0.0454757$}\\
  \hline
 NH, case 2&$\pm 1.59508$&$\mp 0.0647305$&$\pm 3.09629$&\makecell{NH and $\sin^2 \theta_{23}=0.551$ and bfv's\\ for remaining
 oscillation parameters,\\ $(m_{1},m_{2},m_{3})=$\\$(0.0707692,0.0712957,0.0869084)$,\\ $m_{\beta\beta}=0.0712444$}\\
  \hline
   \end{tabular}
  \end{center}
\begin{center}
   \begin{tabular}{|l|r|r|r|}
  \hline
  \multicolumn{1}{|c|}{\sffamily $\delta$}
 &\multicolumn{1}{|c|}{\sffamily $\rho_{1}$}&\multicolumn{1}{|c|}{\sffamily $\rho_{2}$}&\multicolumn{1}{|c|}{\sffamily works with}\\
  \hline
  $\pm 1.56553$&$\pm 0.0733633$&$\pm 3.19198$&\makecell{IH and $\sin^2 \theta_{23}=0.441$ and bfv's\\ for
  remaining oscillation parameters,\\ $(m_{1},m_{2},m_{3})=$\\$(0.0820116,0.0824663,0.065274)$,\\ $m_{\beta\beta}=0.0817407$}\\
  \hline
  \end{tabular}
  \captionof{table}{Results from $P_{4}$ type texture. Masses are given in eVs.}
  \label{tab04}
  \end{center}
\section{ Resonant Leptogenesis}
\la{Resonant}
Expression for $\delta_N(M)$ with effects of $\lambda_{\mu, \tau}$ and ignoring $\lambda_{e}$, is given by Eq. (\ref{eta-vs-eta1}).
The CP asymmetries $\ep_1$ and $\ep_2$
generated by out-of-equilibrium decays of the quasi-degenerate fermionic components of $N_1$ and $N_2$ states respectively are given by
\cite{Pilaftsis:1997jf}, \cite{Pilaftsis:2003gt}:\footnote{In Appendix \ref{app-scalar-asym} we investigate the contribution to the baryon asymmetry
via decays of the scalar components of the RHN superfields. As we show, these effects are less than $3.4\%$.}
\beq
\ep_1=\fr{{\rm Im}[(\hat{Y}_{\nu }^{\dagger}\hat{Y}_{\nu })_{21}]^2}
{(\hat{Y}_{\nu }^{\dagger}\hat{Y}_{\nu })_{11}(\hat{Y}_{\nu }^{\dagger}\hat{Y}_{\nu })_{22}}
\fr{\l M_2^2-M_1^2\r M_1\Ga_2}{\l M_2^2-M_1^2\r^2+M_1^2\Ga_2^2}~,
~~~~~~~~~~~~\ep_2=\ep_1(1\lrar 2)~.
\la{res-lept-asym}
\eeq
Here $M_1, M_2$ (with $M_2>M_1$) are the mass eigenvalues of the RHN  mass matrix. These masses, within our scenario, are given in
(\ref{MN-diag-tion}) with the splitting parameter given in Eq. (\ref{abs-deltaN}).
For the decay widths, here we will use more accurate expressions \cite{Giudice:2003jh}:
\beq
\Gamma_{N_i}=\frac{M_{i}}{8\pi}(\hat{Y^{\dagger}}\hat{Y})_{ii}\Biggl(\left(1-4\frac{M^{2}_{S}}{M^{2}_{i}}\right)^{\frac{1}{2}}+s^{2}_{\beta}+c^{2}_{\beta}\left(1-\frac{M^{2}_{S}}{M^{2}_{i}}\right)^2\Biggr),~~~ \la{gama-n1}
\eeq
\la{gama-n2}
where $M_S$ is the SUSY scale and we assume that all SUSY states have the common mass equal to this scale. $s_{\beta }$ and $c_{\beta}$ are
short hand notations for $\sin \beta$ and $\cos \beta $ respectively. $N_i$ decays proceed via
$N_i\to h_ul_i$ and
$N_i\to  \tilde{h}_u\tilde{l}_i$ channels. Upon derivation of (\ref{gama-n1}) we took into account that $h_u$ is a linear combination of the
SM Higgs doublet $h_{SM}$ and the heavy Higgs  doublet  $H$:
$h_u\simeq s_{\beta }h_{SM}+
c_{\beta}H$. Mass of the $h_{SM}$ has been ignored, while the mass of the $H$ has been taken$\simeq M_S$.
 Moreover, the imaginary part of $[(\hat{Y}_{\nu }^{\dagger}\hat{Y}_{\nu })_{21}]^2$
will be computed with help of (\ref{YdagY-sq}) and (\ref{PYYP-form}) with the relevant phase given in Eq. (\ref{eta-eta1-aprox}).
Using  general expressions (\ref{eta-eta1-aprox}) and (\ref{abs-deltaN}) for the given neutrino model we will compute $\eta - \eta'$ and
$|\delta_N(M)|$.
With these, since we know the possible values of the phase $\phi $ [see Eqs.
(\ref{p1texture}),(\ref{p2texture}),(\ref{p3texture}),(\ref{p4texture})], and with the help of the relations
(\ref{relations-p1}), (\ref{relations-p2}), (\ref{relations-p3}), (\ref{relations-p4}) we can
compute $\ep_{1,2}$ in terms of $|M|$  and $a_2$ or $a_1$ (depending on the texture we are dealing with).
Recalling that the  lepton asymmetry is converted to the baryon asymmetry via sphaleron processes
\cite{Kuzmin:1985mm}, with the relation $\frac{n_{b}^{f}}{s}\simeq -1.48\times10^{-3}({\kappa_f}^{(1)}\epsilon_1+{\kappa_f}^{(2)}\epsilon_2) $
we can compute the baryon asymmetry.  The notion $n_{b}^{f}$ is used for the baryon asymmetry created through the decays of the fermionic components of $N_{1,2}$ superfields.  The net baryon asymmetry $n_b$ receives the contribution from the decays of the scalar components $\tilde N_{1,2}$. The latter contribution we denote by $\tilde {n}_b$. The computation of it (being suppressed in comparison with $n_{b}^{f}$) will be discussed in Appendix \ref{app-scalar-asym}.
For the efficiency factors ${\kappa_f}^{(1,2)}$  we will use the extrapolating expressions  \cite{Giudice:2003jh} (see Eq. (40) in Ref.
\cite{Giudice:2003jh}),
with  ${\kappa_f}^{(1)}$ and ${\kappa_f}^{(2)}$ depending on the mass scales
${\tilde{m}}_1=\frac{v_u^2(M)}{M_1}(\hat{Y}_{\nu}^{\dag}{\hat{Y_{\nu}}})_{11}$ and
${\tilde{m}}_2=\frac{v_u^2(M)}{M_2}(\hat{Y}_{\nu}^{\dag}{\hat{Y_{\nu}}})_{22}$ respectively.

Within our studies we will consider the RHN masses $\simeq |M|\stackrel{<}{_\sim }10^7$~GeV. With this, we will not have the relic gravitino
problem \cite{Khlopov:1984pf}, \cite{Davidson:2002qv}.
For simplicity, we consider all  SUSY particle masses to be equal to  $M_S< |M|$, with $M_S$ identified with the SUSY scale, below
which we have just SM.
As it turns out, via the RG factors, the asymmetry also depends on the top quark mass.

It is remarkable that within some models the observed baryon asymmetry
\beq
\l \! \fr{n_b}{s}\!\r_{\rm exp} =\l 8.65\pm 0.085\r \tm 10^{-11}
\la{nbs-exp}
\eeq
(the recent value reported by WMAP and Planck  \cite{Ade:2015xua}), can be obtained even for low values of the MSSM parameter $\tan
\bt =\fr{v_u}{v_d}$
(defined at the SUSY scale $\mu = M_S$).

Below, we perform analysis for each of these $P_{1,2,3,4}$ cases (and for revised models of Ref.\cite{Babu:2008kp} discussed in Sect.\ref{improvement}) in turn and present our results. As an input for the top's running mass we will use the central value,
while for the SUSY scale $M_S$ we will consider two cases:
$$
m_t(m_t)=163.48~{\rm GeV} ,
$$
\begin{equation}
{\rm Case ~\bf (I)}:~~M_S=10^3~{\rm GeV},~~~~~
{\rm Case ~\bf (II)}:~~
M_S=2\times 10^3~{\rm GeV} .
\label{inp-mt-MS}
\end{equation}

Procedure of our RG calculation and used schemes are described in Appendix \ref{app-baund-match}. As it was shown in \cite{Achelashvili:2016nkr}, for neutrino mass matrix textures $P_{1,2,3,4}$, we will be able to relate the cosmological
phase $\phi $ to the CP violating phase $\delta $. We will introduce the notation:
\beq
\mathcal{A}_{ij}=U^{\ast}_{i1}U^{\ast}_{j1}m_{1}+U^{\ast}_{i2}U^{\ast}_{j2}m_{2}e^{i\rho_{1}}+U^{\ast}_{i3}U^{\ast}_{j3}m_{3}e^{i\rho_{2}},
\la{qq}
\eeq
which will be convenient for writing down expressions for the $\phi $ and for expressing neutrino Dirac type Yukawa couplings in terms of one
independent coupling element. (The latter will be selected by the convenience.)
\\
\\
\textbf{ For $P_1$ Texture}
\\
For this case, using the form of the $M_{\nu}$ [given by Eq. (\ref{M-p1}) and derived within our setup] in the relation  (\ref{nu1}) and equating
appropriate matrix elements of the both sides, we will be able to calculate the phase $\phi $ \cite{Achelashvili:2016nkr},
\cite{Achelashvili:2016trx}:
\\
\beq
\phi ={\rm Arg}\left[\left(\frac{\mathcal{A}_{23}}{\sqrt{\mathcal{A}_{22}\mathcal{A}_{33}}}\mp
\sqrt{\frac{\mathcal{A}^{2}_{23}}{\mathcal{A}_{22}\mathcal{A}_{33}}-1}\right)^{2}\right] \la{p1texture}.
\eeq
Note, all elements at right hand side of Eq. (\ref{p1texture}) are known and therefore the phase $\phi$ is calculable in this
case.\footnote{Same will be true also for textures $P_2, P_3$ and $P_4$.}
Moreover, expressing $a_3, b_{2,3}$ in terms of $a_2$ (taking $a_2$ to be an independent variable) and other known and/or predicted
parameters, we will have:
\beq
a_{3}=\frac{a_2}{r_{\nu
3}}\frac{1}{|\mathcal{A}_{22}|}\Bigg|\mathcal{A}_{23}\pm\sqrt{\mathcal{A}^{2}_{23}-\mathcal{A}_{22}\mathcal{A}_{33}}\Bigg|,~~~b_{2}=\frac{|\mathcal{A}_{22}|}{2|\bar{m}|a_{2}},~~~
b_{3}=\frac{|\mathcal{A}_{33}|}{2|\bar{m}|a_{3}r^{2}_{\nu 3}} .\la{relations-p1}
\eeq
As we see from Eqs. (\ref{p1texture}) and (\ref{relations-p1}), there is a pair of solutions. When for the $a_3$ in (\ref{relations-p1}) we
are taking the $"+"$ sign, in (\ref{p1texture}) we should take the sign $"-"$, and vice versa. (The same applies to the cases of textures $P_{2, 3,
4}$.) For this case, the baryon asymmetry via the resonant leptogenesis has been investigated in Ref. \cite{Achelashvili:2016trx}. In this
work, for the decay widths we use more refined expressions of Eq. (\ref{gama-n1}). Because of this, the values of $\tan \beta $ (given in
Table  \ref{tab8}) are slightly different.
Since in this model $(Y_{\nu })_{31}$ and $(Y_{\nu })_{32}$ are non-zero, according to Eq. (\ref{eta-vs-eta1}) the mismatch $\eta -\eta'$
(e.g. CP asymmetry) is mainly arising due to $\xi_{\tau}$. However, in numerical calculations we have also taken into account the contribution of $\xi_{\mu }$. The results are given in Table \ref{tab8} (for more explanations see also caption of this
table). While in the table we vary the values of $M$ and $\tan \beta $, the cases with {\bf I} and {\bf II} correspond respectively to the cases {\bf (I)} and   {\bf (II)} of
Eq. (\ref{inp-mt-MS}) (i.e. $M_S=1$ and $2$~TeV resp.). For the definition of the RG factors given in this table see Appendix A.2 of Ref. \cite{Achelashvili:2016trx}. For finding maximal
values of the Baryon asymmetries (given in Tab.\ref{tab8}) we have varied the parameter $a_2$.  As we see, the value of the net baryon asymmetry $n_b$ slightly differs from $n_{b}^{f}$. This is due to the contribution from $\tilde {n}_b$ [coming from the right handed sneutrino (RHS) decays], which is small (less than 3.4\% of $n_{b}^{f}$). Details of $\tilde {n}_b$'s calculations are discussed in Appendix \ref{app-scalar-asym}.
\begin{center}
  \begin{tabular}{|l|r|r|r|r|r|c|c|c|}
  \hline
 \multicolumn{1}{|c|}{\sffamily Case}&\multicolumn{1}{|c|}{\sffamily $\!M\mathrm{(GeV)}\!$}&\multicolumn{1}{|c|}{\sffamily
 $\tan\beta$}&\multicolumn{1}{|c|}{\sffamily $r_{\bar{m}}$}&\multicolumn{1}{|c|}{\sffamily $r_{v_{u}}$}&\multicolumn{1}{|c|}{\sffamily
 $\kappa_{N}$}&\multicolumn{1}{|c|}{\sffamily $10^{5}\times\xi_{\tau}$}&\multicolumn{1}{|c|}{\sffamily
 $10^{11}\!\times\!\left(\frac{n_b^f}{s}\right)_{{max}}$}&\multicolumn{1}{|c|}{\sffamily
 $10^{11}\!\times\!\left(\frac{n_b}{s}\right)_{{max}}$}\\
  \hline
\textbf{(I.1)}&$3\cdot10^3$&1.72&0.8868&0.9714 &1.206&6.106 &8.29&8.57\\
  \hline
\textbf{(I.2)}&$10^4$&1.619&0.832&0.9523&1.2322&5.303&8.34&8.6\\
  \hline
\textbf{(I.3)}&$10^5$&1.664&0.7482&0.9203&1.1807&4.821&8.36&8.6\\
  \hline
\textbf{(I.4)}&$10^6$&1.719&0.682&0.8923&1.1345&4.381&8.37&8.6\\
  \hline
\textbf{(I.5)}&$10^7$&1.773&0.6291&0.8676&1.0971&3.937&8.37&8.6\\
  \hline
  \hline
\textbf{(II.1)}&$6\cdot10^3$&1.701&0.8689&0.9678&1.175&5.897&8.294&8.57\\
  \hline
\textbf{(II.2)}&$10^4$&1.615&0.8464&0.9599&1.1994&5.365&8.334&8.59\\
  \hline
\textbf{(II.3)}&$10^5$&1.625&0.7629&0.9283&1.1669&4.755&8.36&8.6\\
  \hline
\textbf{(II.4)}&$10^6$&1.678&0.6974&0.9008&1.1243&4.321&8.36&8.6\\
  \hline
\textbf{(II.5)}&$10^7$&1.731&0.645&0.8765&1.0894&3.887&8.36&8.6\\
  \hline
  \end{tabular}
  \captionof{table}{Texture $P_1$, normal hierarchy: Baryon asymmetry for various values of $M$ and for minimal
(allowed) value of $\tan\beta$. With neutrino oscillation parameters
and results given in the Table \ref{tab01} and computed from Eq. (\ref{p1texture}) $\phi =\pm 1.264$. For all cases
$r_{\nu 3}\simeq 1$.}
  \label{tab8}
\end{center}
\textbf{ For $P_2$ Texture}
\\
With a pretty similar procedure, for this case we get:
\beq
\phi ={\rm Arg}\left[\left(\frac{\mathcal{A}_{23}}{\sqrt{\mathcal{A}_{22}\mathcal{A}_{33}}}\mp
\sqrt{\frac{\mathcal{A}^{2}_{23}}{\mathcal{A}_{22}\mathcal{A}_{33}}-1}\right)^{2}\right] \la{p2texture}.
\eeq
Expressing $a_3, b_{2,3}$ in terms of $a_2$ and other parameters (yet known or predicted in this scenario), we will have:
\beq
a_{3}=\frac{a_2}{r_{\nu
3}}\frac{1}{|\mathcal{A}_{22}|}\Bigg|\mathcal{A}_{23}\pm\sqrt{\mathcal{A}^{2}_{23}-\mathcal{A}_{22}\mathcal{A}_{33}}\Bigg|,~~~b_{2}=\frac{|\mathcal{A}_{22}|}{2|\bar{m}|a_{2}},~~~
b_{3}=\frac{|\mathcal{A}_{33}|}{2|\bar{m}|a_{3}r^{2}_{\nu 3}} \la{relations-p2}
\eeq
Results for this case are presented in Table \ref{tab5}.
\begin{center}
  \begin{tabular}{|l|r|r|r|r|r|c|c|c|}
  \hline
 \multicolumn{1}{|c|}{\sffamily Case}&\multicolumn{1}{|c|}{\sffamily $\!M\mathrm{(GeV)}\!$}&\multicolumn{1}{|c|}{\sffamily
 $\tan\beta$}&\multicolumn{1}{|c|}{\sffamily $r_{\bar{m}}$}&\multicolumn{1}{|c|}{\sffamily $r_{v_{u}}$}&\multicolumn{1}{|c|}{\sffamily
 $\kappa_{N}$}&\multicolumn{1}{|c|}{\sffamily $10^{5}\times\xi_{\tau}$}&\multicolumn{1}{|c|}{\sffamily
 $10^{11}\!\times\!\left(\frac{n_b^f}{s}\right)_{{max}}$}&\multicolumn{1}{|c|}{\sffamily
 $10^{11}\!\times\!\left(\frac{n_b}{s}\right)_{{max}}$}\\
  \hline
\textbf{(I.1)}&$3\cdot10^3$&1.948&0.8908 &0.9725&1.1439&7.264 & 8.306&8.57\\
  \hline
\textbf{(I.2)}&$10^4$&1.833&0.8412&0.955&1.1543 &6.242&8.35&8.6\\
  \hline
\textbf{(I.3)}&$10^5$&1.881&0.7647&0.9254&1.1158 &5.692&8.37&8.6\\
  \hline
\textbf{(I.4)}&$10^6$&1.938&0.7039&0.8994&1.0821&5.182&8.36&8.6\\
  \hline
\textbf{(I.5)}&$10^7$&1.996&0.6554&0.8766&1.0544 &4.671&8.36&8.6\\
  \hline
  \hline
\textbf{(II.1)}&$6\cdot10^3$&1.933&0.8728&0.9689&1.1201&7.058&8.314&8.57\\
  \hline
\textbf{(II.2)}&$10^4$&1.836&0.8526&0.9616&1.133 &6.373 &8.35&8.6\\
  \hline
\textbf{(II.3)}&$10^5$&1.843&0.7771&0.9326 &1.1063&5.638&8.36&8.6\\
  \hline
\textbf{(II.4)}&$10^6$&1.9&0.7175&0.9072&1.0748 &5.14&8.37&8.6\\
  \hline
\textbf{(II.5)}&$10^7$&1.956&0.6697&0.8848&1.049&4.632&8.37&8.6\\
  \hline
  \end{tabular}
 \captionof{table}{Texture $P_2$, normal hierarchy: Baryon asymmetry for various values of $M$ and for minimal
(allowed) value of $\tan\beta$. With neutrino oscillation parameters
and results given in the Table  \ref{tab02} and computed from Eq. (\ref{p2texture})  $\phi =\pm 1.1$. For all cases
$r_{\nu 3}\simeq 1$.}
  \label{tab5}
\end{center}
\textbf{ For $P_3$ Texture}
\begin{center}
  \begin{tabular}{|l|r|r|r|r|r|c|c|c|}
  \hline
 \multicolumn{1}{|c|}{\sffamily Case}&\multicolumn{1}{|c|}{\sffamily $\!M\mathrm{(GeV)}\!$}&\multicolumn{1}{|c|}{\sffamily
 $\tan\beta$}&\multicolumn{1}{|c|}{\sffamily $r_{\bar{m}}$}&\multicolumn{1}{|c|}{\sffamily $r_{v_{u}}$}&\multicolumn{1}{|c|}{\sffamily
 $\kappa_{N}$}&\multicolumn{1}{|c|}{\sffamily $10^{5}\times\xi_{\tau}$}&\multicolumn{1}{|c|}{\sffamily
 $10^{11}\!\times\!\left(\frac{n_b^f}{s}\right)_{{max}}$}&\multicolumn{1}{|c|}{\sffamily
 $10^{11}\!\times\!\left(\frac{n_b}{s}\right)_{{max}}$}\\
  \hline
\textbf{(I.1)}&$3\cdot10^3$&7.158&0.904 &0.9761&1.0076&76.29 &8.49&8.59\\
  \hline
\textbf{(I.2)}&$10^4$&6.802&0.8717&0.9635&0.9983 &64.79&8.508&8.6\\
  \hline
\textbf{(I.3)}&$10^5$&6.922&0.82&0.9417&0.9819  &59.11&8.51&8.6\\
  \hline
\textbf{(I.4)}&$10^6$&7.074&0.7789&0.9225&0.9692&53.92&8.51&8.6\\
  \hline
\textbf{(I.5)}&$10^7$&7.227&0.7467 &0.9056 &0.96  &48.65&8.51&8.6\\
  \hline
  \hline
\textbf{(II.1)}&$6\cdot10^3$&7.146&0.8852&0.9723 &0.9986&75.06&8.5&8.6\\
  \hline
\textbf{(II.2)}&$10^4$&6.85&0.8725 &0.9672&0.9954 &67.24 &8.5&8.6\\
  \hline
\textbf{(II.3)}&$10^5$&6.858&0.8229&0.946 &0.9802 &59.44&8.51&8.6\\
  \hline
\textbf{(II.4)}&$10^6$&7.003&0.7835&0.9274&0.9684  &54.17 &8.51&8.6\\
  \hline
\textbf{(II.5)}&$10^7$&7.151&0.7524&0.9109&0.9597&48.87&8.51&8.6\\
  \hline
  \end{tabular}
  \captionof{table}{Texture $P_3$, normal hierarchy: Baryon asymmetry for various values of $M$ and for minimal
(allowed) value of $\tan\beta$. With neutrino oscillation parameters
and results given in the Table \ref{tab03} and computed from Eq. (\ref{p3texture}) (for NH case) $\phi =\pm 2.92$. For all cases
$r_{\nu 3}\simeq 1$.}
  \label{tab6}
\end{center}
\begin{center}
  \begin{tabular}{|l|r|r|r|r|r|c|c|c|}
  \hline
 \multicolumn{1}{|c|}{\sffamily Case}&\multicolumn{1}{|c|}{\sffamily $\!M\mathrm{(GeV)}\!$}&\multicolumn{1}{|c|}{\sffamily
 $\tan\beta$}&\multicolumn{1}{|c|}{\sffamily $r_{\bar{m}}$}&\multicolumn{1}{|c|}{\sffamily $r_{v_{u}}$}&\multicolumn{1}{|c|}{\sffamily
 $\kappa_{N}$}&\multicolumn{1}{|c|}{\sffamily $10^{5}\times\xi_{\tau}$}&\multicolumn{1}{|c|}{\sffamily
 $10^{11}\!\times\!\left(\frac{n_b^f}{s}\right)_{{max}}$}&\multicolumn{1}{|c|}{\sffamily
 $10^{11}\!\times\!\left(\frac{n_b}{s}\right)_{{max}}$}\\
  \hline
\textbf{(I.1)}&$3\cdot10^3$&27.11&0.905 &0.9764 &1.0038&1154.3 &8.515&8.6\\
  \hline
\textbf{(I.2)}&$10^4$&25.824&0.8738 &0.9641&0.9938 &980.4&8.52&8.6\\
  \hline
\textbf{(I.3)}&$10^5$&26.138&0.8234&0.9427&0.9784  &894.7&8.53&8.6\\
  \hline
\textbf{(I.4)}&$10^6$&26.55&0.7833&0.9238&0.9667&815.9&8.53&8.6\\
  \hline
\textbf{(I.5)}&$10^7$&26.96&0.7515 &0.9071 &0.9583&736&8.53&8.6\\
  \hline
  \hline
\textbf{(II.1)}&$6\cdot10^3$&27.1&0.886&0.9725 &0.995 &1135.1&8.516&8.6\\
  \hline
\textbf{(II.2)}&$10^4$&26.061&0.8739 &0.9676&0.991 &1017.9 &8.518&8.6\\
  \hline
\textbf{(II.3)}&$10^5$&25.979&0.8259 &0.9469  &0.9766 &899.4&8.52&8.6\\
  \hline
\textbf{(II.4)}&$10^6$&26.38&0.7875&0.9285&0.9657  &819.9 &8.53&8.6\\
  \hline
\textbf{(II.5)}&$10^7$&26.783&0.757&0.9123&0.9578 &739.6 &8.53&8.6\\
  \hline
  \end{tabular}
  \captionof{table}{Texture $P_3$, inverted hierarchy: Baryon asymmetry for various values of $M$ and for minimal
(allowed) value of $\tan\beta$. With neutrino oscillation parameters
and results given in the Table \ref{tab03} and computed from Eq. (\ref{p3texture}) (for IH case) $\phi =\pm 3.124$. For all cases
$r_{\nu 3}\simeq 1$.}
  \label{tab7}
\end{center}
(For notations and definitions see also Appendix \ref{app-nuRG} of the present paper.)
\beq
 \quad \phi ={\rm Arg}\left[\left(\frac{\mathcal{A}_{13}}{\sqrt{\mathcal{A}_{11}\mathcal{A}_{33}}}\mp
 \sqrt{\frac{\mathcal{A}^{2}_{13}}{\mathcal{A}_{11}\mathcal{A}_{33}}-1}\right)^{2}\right] \la{p3texture}.
 \eeq
 Expressing $a_3, b_{1,3}$ in terms of $a_1$ and other fixed parameters, we will have:
 \beq
 a_{3}=\frac{a_1}{r_{\nu
 3}}\frac{1}{|\mathcal{A}_{11}|}\Bigg|\mathcal{A}_{13}\pm\sqrt{\mathcal{A}^{2}_{13}-\mathcal{A}_{11}\mathcal{A}_{33}}\Bigg|,~~~b_{1}=\frac{|\mathcal{A}_{11}|}{2|\bar{m}|a_{1}},~~~
 b_{3}=\frac{|\mathcal{A}_{33}|}{2|\bar{m}|a_{3}r^{2}_{\nu 3}} \la{relations-p3}
 \eeq
 Results for this texture for cases of NH and IH neutrinos are presented in Tables \ref{tab6} and \ref{tab7} respectively.
 \\
 \\
\textbf{ For $P_4$ Texture}
\\
For this case cosmological phase is given by:
\beq
 \quad \phi ={\rm Arg}\left[\left(\frac{\mathcal{A}_{12}}{\sqrt{\mathcal{A}_{11}\mathcal{A}_{22}}}\mp
 \sqrt{\frac{\mathcal{A}^{2}_{12}}{\mathcal{A}_{11}\mathcal{A}_{22}}-1}\right)^{2}\right] \la{p4texture}.
 \eeq
 Expressing $a_1, b_{1,2}$ in terms of $a_2$ and other known and/or predicted parameters, we will have:
 \beq
 a_{1}=\frac{|\mathcal{A}_{11}|}{|\mathcal{A}_{12}\pm\sqrt{\mathcal{A}^{2}_{12}-\mathcal{A}_{11}\mathcal{A}_{22}}|}a_{2},~~~b_{1}=\frac{|\mathcal{A}_{11}|}{2|\bar{m}|a_{1}},~~~
 b_{2}=\frac{|\mathcal{A}_{22}|}{2|\bar{m}|a_{2}} \la{relations-p4}
 \eeq
 In this scenario, since
 $(Y_{\nu })_{31}$ and $(Y_{\nu })_{32}$ are zero, according to Eq. (\ref{eta-vs-eta1}) the mismatch $\eta -\eta'$ (e.g. CP asymmetry) is
 arising due to $\xi_{\mu }$. Since the latter is suppressed by $\lambda_{\mu }^2$, as it turns out  large values of the $\tan \beta $ are
 required and only in NH case needed amount of the Baryon asymmetry can be generated.
Results are given in Table \ref{tab9}.
 \\
\begin{center}
  \begin{tabular}{|l|r|r|r|r|r|c|c|c|}
  \hline
 \multicolumn{1}{|c|}{\sffamily Case}&\multicolumn{1}{|c|}{\sffamily $\!M\mathrm{(GeV)}\!$}&\multicolumn{1}{|c|}{\sffamily
 $\tan\beta$}&\multicolumn{1}{|c|}{\sffamily $r_{\bar{m}}$}&\multicolumn{1}{|c|}{\sffamily $r_{v_{u}}$}&\multicolumn{1}{|c|}{\sffamily
 $\kappa_{N}$}&\multicolumn{1}{|c|}{\sffamily $10^{4}\times\xi_{\mu}$}&\multicolumn{1}{|c|}{\sffamily
 $10^{11}\!\times\!\left(\frac{n_b^f}{s}\right)_{{max}}$}&\multicolumn{1}{|c|}{\sffamily
 $10^{11}\!\times\!\left(\frac{n_b}{s}\right)_{{max}}$}\\
  \hline
\textbf{(I.1)}&$3\cdot10^3$&64.639&0.9048 &0.9763&1.0349&3.111 &8.518&8.6\\
  \hline
\textbf{(I.2)}&$10^4$&62.213&0.873&0.9638&1.0212 &2.638&8.52&8.6\\
  \hline
\textbf{(I.3)}&$10^5$&62.02&0.8203&0.9418&1.0059 &2.416&8.53&8.6\\
  \hline
\textbf{(I.4)}&$10^6$&62.006&0.7767&0.9218 &0.994&2.213&8.53&8.6\\
  \hline
\textbf{(I.5)}&$10^7$&62&0.7404&0.9037&0.9848  &2.008&8.53&8.6\\
  \hline
  \hline
\textbf{(II.1)}&$6\cdot10^3$&65.28&0.8859&0.9725&1.0208 &3.045&8.517&8.59\\
  \hline
\textbf{(II.2)}&$10^4$&63.398&0.8735&0.9675&1.0145 &2.728 &8.525&8.59\\
  \hline
\textbf{(II.3)}&$10^5$&62.548&0.8239&0.9463 &0.9996 &2.417&8.53&8.6\\
  \hline
\textbf{(II.4)}&$10^6$&62.528&0.7827&0.9271&0.9886 &2.211&8.53&8.6\\
  \hline
\textbf{(II.5)}&$10^7$&62.535&0.7484&0.9097&0.9803&2.005&8.53&8.6\\
  \hline
  \end{tabular}
  \captionof{table}{Texture $P_4$, normal hierarchy: Baryon asymmetry for various values of $M$ and for minimal (allowed) value of
  $\tan\beta$. With neutrino oscillation parameters
and results given in the Table \ref{tab04}, NH, case 1, and $\phi $ computed from Eq. (\ref{p4texture}) (for NH case)  $\phi =\pm 2.872$.}
  \label{tab9}
\end{center}
\section{Revising Textures of Ref. \cite{Babu:2008kp} and Improved Versions}
\la{improvement}

In this section we revise the textures considered in the work \cite{Babu:2008kp}. Since some of them are excluded by
the current neutrino data \cite{recent-nu-data}(see also Eq. (\ref{ever-bfv})), we apply $d=5$ contributions (in a spirit of section \ref{RG-CP-1})
and achieve their compatibility with the best fit values. Together with this, we investigate resonant leptogenesis and show that
one loop corrections via $\lam_{\tau}$ and/or $\lam_{\mu }$ are crucial. In \cite{Babu:2008kp}, while ignoring $\lambda_{\mu}$ the two loop correction to $\lambda_{\tau}$ was taken into account and this suggested for textures A and B$_1$ specific low bounds on the values of $\tan \beta $.
As demonstrated below, one loop effects of $\lam_{\tau }$ (giving dominant contribution for textures A and B$_1$) and $\lam_{\mu }$ (for the texture B$_2$)
significantly change results.

In the setup of two degenerate RHNs, in Ref. \cite{Babu:2008kp} the following three possible one texture zero  neutrino Dirac Yukawa couplings
have been considered :
\beq
{\rm Texture~ A:}~~~~~~~~~Y_{\nu}=\l \!\begin{array}{cc}
a_1e^{i\alpha_1}&0\\
a_2e^{i\alpha_2}&b_2e^{i\beta_2}\\
a_3e^{i\alpha_3}&b_3e^{i\beta_3}
  \end{array}\!\r  ~,
\la{textureA}
\eeq
\beq
{\rm Texture~ B_1:}~~Y_{\nu}=\l \!\begin{array}{cc}
a_1e^{i\alpha_1} &b_1e^{i\beta_1}\\
a_2e^{i\alpha_2} &0\\
 a_3e^{i\alpha_3} &b_3e^{i\beta_3}
  \end{array}\!\r  ,~~~~~~~
{\rm Texture~ B_2:}~~Y_{\nu}=\l \!\begin{array}{cc}
a_1e^{i\alpha_1} &b_1e^{i\beta_1} \\
a_2e^{i\alpha_2} &b_2e^{i\beta_2} \\
a_3e^{i\alpha_3} &0
  \end{array}\!\r,
\la{textureB12}
\eeq
where for notational consistency with the whole paper, we have shown phases $\al_i, \bt_j$, while assuming that the couplings
$a_i, b_j$ are real.\footnote{On the contrary, in Ref. \cite{Babu:2008kp}, without writing down the phase factors, $a_i$ and $b_j$ were treated
as
a complex parameters.}
Below we will (re)investigate these textures in turn.

\vspace{0.3cm}

{\bf Texture A}


The A Yukawa texture can be written as:
 \beq
{\rm Texture ~A}:~~Y_{\nu }=\begin{pmatrix}
a_1e^{i\alpha_1}&0\\
a_2e^{i\alpha_2}&b_2e^{i\beta_2}\\
a_3e^{i\alpha_3}&b_3e^{i\beta_3}
\end{pmatrix}
=
\begin{pmatrix}
e^{ix} & 0&0\\
0 & e^{iy}&0\\
0 & 0&e^{iz}
\end{pmatrix}
\begin{pmatrix}
\hspace{-0.4cm}a_{1} & 0\\
\hspace{-0.4cm}a_{2} &  b_{2}\\
a_3e^{i\phi} &b_3
\end{pmatrix}
\begin{pmatrix}
e^{i\omega} & 0\\
0 & e^{i\rho}
\end{pmatrix}, \nonumber
\eeq
\beq
{\rm with}   ~~~x= \alpha_1 -\alpha_2+\beta_2-\rho, \quad
y= \beta_2-\rho, \quad  z= \beta_3-\rho, \quad
\omega=\alpha_2-\beta_2+\rho, \quad \phi=\alpha_3-\alpha_2.
\la{YA}
\eeq
As we see, besides the phase $\phi $ all phases are factored out and have no physical relevance.
With the RHN mass matrix of Eq.(\ref{M-prime-N}), via the see-saw[see expression in Eq.(\ref{seesaw})] we will get the  light neutrino mass
matrix:
\beq
M_{\nu}^{({\rm A})}(M_Z)=\l \!\begin{array}{ccc}
0&a_1b_2&a_1b_3r_{\nu 3}\\
a_1b_2&2a_2b_2&(a_2b_3+a_3b_2e^{i\phi })r_{\nu 3}\\
a_1b_3r_{\nu 3}& (a_2b_3+a_3b_2e^{i\phi })r_{\nu 3} &2a_3b_3e^{i\phi }r_{\nu 3}^2\end{array}\!\r \!\bar m~,
\la{Mnu-A}
\eeq
[For definitions of $\bar m$, $r_{\nu 3}$ and proper explanations see respectively Eq. (\ref{bar-m-rbar}) and also Eqs. (A.17), (A.18) of Ref.
\cite{Achelashvili:2016trx}, and comments therein.]
This neutrino mass texture has only two non-zero mass eigenvalues.
As it was shown in \cite{Babu:2008kp}, this for NH ($m_1=0$) and IH ($m_3=0$) neutrino mass patterns, gives respectively the predictive
relations
$\tan \te_{13}=\sqrt{\fr{m_2}{m_3}}s_{12}$ and $\tan \te_{12}=\sqrt{\fr{m_1}{m_2}}$.
Both of them are in a gross conflict with the current neutrino data, which
exclude this scenario.

\vspace{0.3cm}

{\bf ${\rm A}'$ Neutrino Texture: Improved Version}


The drawbacks coming from the  A neutrino mass matrix (\ref{Mnu-A}) can be avoided by adding $d_5$ term to one of the entries.
Here we consider this addition to
the $(2,3)$ and $(3,2)$ elements of the light neutrino mass matrix, which would make the model viable. (We refer to this improved version
of (\ref{Mnu-A}) as the ${\rm A}'$ neutrino texture.)
 After this, the $M_{\nu }$ will have the form:
\beq
M_{\nu}^{({\rm A}')}(M_Z)=\l \!\begin{array}{ccc}
0&a_1b_2&a_1b_3r_{\nu 3}\\
a_1b_2&2a_2b_2&(a_2b_3\!+\!a_3b_2e^{i\phi })r_{\nu 3}\!+\!d_5\\
a_1b_3r_{\nu 3}& (a_2b_3\!+\!a_3b_2e^{i\phi })r_{\nu 3}\!+\!d_5 &2a_3b_3e^{i\phi }r_{\nu 3}^2\end{array}\!\r \!\bar m .
\la{Mnu-Apr}
\eeq
With this modification, all masses are non-zero. One can check out, that with the fixed phase redefinitions [given in Eq. (\ref{YA})], in
general $d_5$ is a complex parameter. Thus, together with additional mass, we will have
one more independent phase.
As it turns out, only NH scenario is possible to realize.
 Therefore
 as additional independent parameters
 we take one of the mass and $\Delta \rho =\rho_1 -\rho_2 $. From the condition $M_{\nu}^{(1,1)}=0$
we have:
\beq
\cos(2\de \!-\!\Delta \rho )\!=\!\fr{m_1^2c_{12}^4\!-\!m_2^2s_{12}^4\!-\!m_3^2t_{13}^4}{2m_2m_3s_{12}^2t_{13}^2} ,~~
\rho_1\!=\!\pi \!-\!{\rm Arg}\lq \fr{m_2}{m_3}s^{2}_{12}+\!t_{13}^2e^{i(2\de -\Delta \rho )}\rq
~~{\rm with}~~\Delta \rho =\rho_1 -\rho_2 .
\la{Apr}
\eeq
(Here and below we use shorthanded notations $t_{ij}\equiv \tan \theta_{ij}$.) From the first relation of (\ref{Apr}) one can check that IH
scenario can not be realized. As far as the NH scenario is concerned, it will work
with low bound on the lightest neutrino mass $m_1$. In fact, the first relation of (\ref{Apr}) gives the allowed range for $m_1$. For
example,
with bfv's of the oscillation parameters (\ref{ever-bfv}) we have:
\beq
0.00239~{\rm eV} \stackrel{<}{_\sim }m_1\stackrel{<}{_\sim }0.00641~{\rm eV}.
\la{m1-range-Apr}
\eeq
Thus, as independent parameters we will take $m_1$ and $\Delta \rho $. We will select them in such a way as to get desirable baryon
asymmetry.
For example, with the choice
\beq
m_1=0.005719~{\rm eV} ,~~~\Delta \rho =4.987
\la{choice-Apr}
\eeq
and bfv's of all measured oscillation parameters with help of (\ref{nh1}) and (\ref{Apr})
for neutrino masses and phases we are getting:
\beqs
\l m_1, m_2, m_3\r \simeq (0.005719,~0.01037,~0.05077)~{\rm eV},
\eeqs
\beq
 ~~~\l \de , \rho_1, \rho_2 \r \simeq \l 2.9639,~2.911 ,~-2.076 \r .
\la{mi-phases-Apr}
\eeq

%
\begin{table}
\vs{0.3cm}
 $$\begin{array}{|c|c|c|c|c|c|c|c|c|}
\hline
{\rm Case}  & {\rm M(GeV}) & \tan \bt  &  r_{\bar m} & r_{v_u} & \ka_N &
 10^{4}\!\tm \! \xi_{\tau } & 10^{11}\!\tm \!\l \!\fr{n_b^f}{s}\!\r_{\rm max} & 10^{11}\!\tm \!\l \!\fr{n_b}{s}\!\r_{\rm max} \\
\hline
 {\bf (I.1)} &  3\cdot 10^3 &1.939 & 0.8907 & 0.9725 &1.1457 & 0.7215 &8.53 &  8.6\\
 \hline
 {\bf (I.2)} &  10^4 &1.838   &0.8414  &0.955  &1.153  &0.6266  &8.53 &8.59  \\
\hline
 {\bf (I.3)} &  10^5 &1.904   &0.7662  &0.9258  &1.111  &0.5793  & 8.53&8.59  \\
 \hline
 {\bf (I.4)} &  10^6 &1.986   &0.7078  & 0.9006 &1.0742  & 0.5374 &8.54 &8.6 \\
 \hline
 {\bf (I.5)} &  10^7 &2.075   &0.6628 & 0.879 & 1.0442 & 0.4956 &8.55 &8.61 \\
 \hline
 \hline
 {\bf (II.1)} &  6\cdot 10^3 &1.928 & 0.8727 & 0.9688 & 1.121 &0.7031 &8.53 & 8.6\\
 \hline
 {\bf (II.2)} &  10^4 &1.84  &0.8527  &0.9617  &1.1322  &0.6393  &8.54 &8.6  \\
\hline
 {\bf (II.3)} &  10^5 &1.869   &0.7784  &0.933  &1.1013  &0.5753  &8.54 &8.6  \\
 \hline
 {\bf (II.4)} &  10^6 &1.949   &0.721  & 0.9083 &1.0672  & 0.5337 &8.54 &8.6 \\
 \hline
 {\bf (II.5)} &  10^7 &2.036  &0.6766 &0.887 &1.0393 & 0.4923 & 8.54&8.6 \\
 \hline
\end{array}$$
\vs{-0.5cm}
\caption{${\rm A}'$ Neutrino Texture, NH. Baryon asymmetry for various values of $M$ and for corresponding minimal (allowed) values of $\tan \bt
$.
With the choice given in Eqs. (\ref{choice-Apr}), (\ref{mi-phases-Apr}) and bfv's of  $s^2_{ij}$. For all cases $r_{\nu 3}\simeq 1$.}
 \vs{-0.3cm}
 \label{nb-Apr}
\end{table}
%
%
%

As far as the baryon asymmetry is concerned, using (\ref{Mnu-Apr}) in (\ref{nu1}) for the CP phase $\phi$ and expressing
couplings $a_{1,3}, b_{2,3}$ in terms of $a_2$
we get
$$
\phi ={\rm Arg}\l \fr{{\cal A}_{12}^2{\cal A}_{33}}{{\cal A}_{13}^2{\cal A}_{22}}\r ,
$$
\beq
a_1=2\left |\fr{{\cal A}_{12}}{{\cal A}_{22}} \right |a_2 ,~~
a_3=\fr{1}{r_{\nu 3}}\left |\fr{{\cal A}_{12}{\cal A}_{33}}{{\cal A}_{22}{\cal A}_{13}}\right |a_2,~~b_2=\fr{|{\cal A}_{22}|}{2|\bar m|a_2},
~~
b_3=\left |\fr{{\cal A}_{13}{\cal A}_{22}}{{\cal A}_{12}} \right |\fr{1}{2r_{\nu 3}|\bar m|a_2}~.
\la{phi-yuk-Apr}
\eeq
For the values of (\ref{choice-Apr}), (\ref{mi-phases-Apr}) and bfv's of  $s^2_{12, 23, 13}$ we get
\beq
\phi =-2.9297~.
\la{phi-Apr-1}
\eeq
With these, and for given values of $M$ and $\tan \bt $ by varying $a_2$ we can investigate the baryon asymmetry.
Results are given in Tab. \ref{nb-Apr}.

\vspace{0.3cm}

{\bf Texture B$_1$}

\vspace{0.3cm}

The B$_1$ Yukawa texture can be written as:
 \beq
{\rm Texture~ B_1}:~~Y_{\nu }=\begin{pmatrix}
a_1e^{i\alpha_1}&b_1e^{i\beta_1}\\
a_2e^{i\alpha_2}&0\\
a_3e^{i\alpha_3}&b_3e^{i\beta_3}
\end{pmatrix}
=
\begin{pmatrix}
e^{ix} & 0&0\\
0 & e^{iy}&0\\
0 & 0&e^{iz}
\end{pmatrix}
\begin{pmatrix}
\hspace{-0.4cm}a_{1} & b_1\\
\hspace{-0.4cm}a_{2} & 0 \\
a_3e^{i\phi} &b_3
\end{pmatrix}
\begin{pmatrix}
e^{i\omega} & 0\\
0 & e^{i\rho}
\end{pmatrix}, \nonumber
\eeq
\beq
{\rm with}   ~~~x= \beta_1 -\rho, \quad
y= \alpha_2-\alpha_1+\beta_1-\rho, \quad  z= \beta_3-\rho, \quad
\omega=\alpha_1-\beta_1+\rho, \quad \phi=\alpha_3-\beta_3-\alpha_1+\beta_1.
\la{YAB1}
\eeq
With the RHN mass matrix of Eq. (\ref{M-prime-N}), via the see-saw we will get the  light neutrino mass matrix:

\beq
M_{\nu}^{({\rm B_1})}(M_Z)=\l \!\begin{array}{ccc}
2a_1b_1&a_2b_1&(a_1b_3+a_3b_1e^{i\phi})r_{\nu 3}\\
a_2b_1&0&a_2b_3r_{\nu 3}\\
(a_1b_3+a_3b_1e^{i\phi})r_{\nu 3}&a_2b_3r_{\nu 3} &2a_3b_3e^{i\phi }r_{\nu 3}^2\end{array}\!\r \!\bar m~,
\la{Mnu-AB1}
\eeq
\\
This neutrino mass texture (referred as B$_1$ neutrino texture) works only for inverted neutrino mass ordering \cite{Babu:2008kp} (with $m_3=0$) and
has two  predictive relations. In particular, in terms of measured oscillation parameters we can calculate
the phases $\de $ and $\rho_1$.
The exact expressions are:
\beq
\cos\delta=\!\fr{m_2(1+t^{2}_{23}t^{2}_{12}s^{2}_{13})-m_1(t^{2}_{12}+t^{2}_{23}s^{2}_{13})}{2t_{23}t_{12}s_{13}(m_1+m_2)} ,~~
\rho_1\!=\!\pi \!-\!{\rm Arg}\lq \fr{(1-t_{23}t_{12}s_{13}e^{-i\delta})^2}{(t_{12}+t_{23}s_{13}e^{-i\delta})^2}\rq
~~ .
\la{AprB1m3zero}
\eeq
\beq
{\rm with}~~~~~m_1=\sqrt{\Delta m_{atm}^{2}-\Delta m_{sol}^{2}} ,~~~~~~m_2=\sqrt{\Delta m_{atm}^{2}} ,~~~~~~m_3=0.
\la{pred-B1m3zero}
\eeq
\\
Although the first expression in (\ref{AprB1m3zero}) excludes the possibility of using the best fit values for all oscillation parameters, it
allows for keeping values of $s^2_{23}$ and $s^2_{13}$  within 1$\sigma$, while confining $s^2_{12}$ to 2$\sigma$.  Remarkably,
needed baryon asymmetry can be achieved with relatively low values of $\tan\beta$. For example,
\\
\beqs
\mathrm{for~ IH~of~ the~ B_1~neutrino~ texture,~~ with}: s_{23}^2=0.604 ~(1\sigma),~~s^{2}_{12}=0.33 ~(2\sigma),~~ s^{2}_{13}=0.023 ~(1\sigma)
\eeqs
\beq
 \Longrightarrow \delta =\pm 0.307,~~\rho_{1}=\pi\mp 0.2192,~~\phi=\pm 3.129
\eeq
\\
($\Delta m^2_{sol}$ and $\Delta m^2_{atm}$ are taken bfv's.) to generate baryon asymmetry of desired amount [$\l \fr{n_b}{s}\r_{max}\simeq
8.59\tm 10^{-11}$] in case of $M=3\cdot 10^3$~GeV and $M_S=1$~TeV the value $\tan\beta=6.32$ is required.

\vs{0.3cm}

{\bf ${\rm B_1}'$ Neutrino Texture: Improved Version}

\vspace{0.3cm}

By addition of the $d_5$ term to (1,3) and (3,1) entries of the B$_1$ neutrino texture (\ref{Mnu-AB1}), the light neutrino mass matrix becomes:
\\
\beq
M_{\nu}^{({\rm B_1}')}(M_Z)=\l \!\begin{array}{ccc}
2a_1b_1&a_2b_1&(a_1b_3+a_3b_1e^{i\phi})r_{\nu 3}+d_{5}\\
a_2b_1&0&a_2b_3r_{\nu 3}\\
(a_1b_3+a_3b_1e^{i\phi})r_{\nu 3}+d_{5}&a_2b_3r_{\nu 3} &2a_3b_3e^{i\phi }r_{\nu 3}^2\end{array}\!\r \!\bar m~,
\la{Mnu-AprB1}
\eeq
\\
which gives all neutrinos massive and opens up a possibility of choosing two variables such as  $m_3$ and $\Delta\rho\equiv\rho_1 -\rho_2$ as
independent ones to operate with. We refer to this (\ref{Mnu-AprB1}) improved version as
the ${\rm B_1}'$ neutrino texture.
From the condition $M_{\nu}^{(2,2)}=0$
we have:
\beq
m_{1}|U_{21}|^{2}=|m_{2}(U_{22})^2+m_{3}(U_{23})^2e^{i\Delta \rho}|,~~
\rho_1\!=\!\pi \!-\!{\rm Arg}\lq \fr{m_{1}(U_{21})^2}{m_2(U_{22})^2+m_3(U_{23})^2e^{i\Delta \rho}}\rq,~~~{\rm with}~~~\Delta \rho =\rho_1
-\rho_2.
~~
\la{AprB1}
\eeq
\\
Out of the numerous values  $\Delta \rho$ and $m_3$ can take on, we select those that are not in conflict with the observed oscillation data
and at the same time together with the minimal allowed value of $\tan\beta$ generate baryon asymmetry of the needed amount.  In case of
Inverted Hierarchy both of these requirements can be satisfied. In particular:
\\
\beq
\mathrm{for~ IH ~of~ the~ {B_1}' ~neutrino ~ texture}: ~~~ m_3=0.00250717~{\rm eV}~~~~ {\rm and}~~~~ \Delta\rho=3.6599  \la{m3delroB1IH}
\eeq
\\
determine numerical values of the rest of masses, phases and eventually the neutrino double beta decay parameter:
\beqs
\l m_1, m_2, m_3\r =(0.049714, ~0.050461, ~0.00250717)~{\rm eV},
\eeqs
\beq
 ~~~\l \de , \rho_1, \rho_2 \r =\l 0.17303,~ 2.9456,~ -0.71436\r .
\la{mi-phasesB1IH}
\eeq
\beq
m_{\bt \bt}\simeq 0.019~{\rm eV}.
\eeq
As far as the baryon asymmetry is concerned, using (\ref{Mnu-AprB1}) in (\ref{nu1}), we get:
\\
\beqs
\phi ={\rm Arg}\l \fr{{\cal A}_{12}^2{\cal A}_{33}}{{\cal A}_{23}^2{\cal A}_{11}}\r ,
\eeqs
\beq
a_1=\frac{1}{2}\left |\fr{{\cal A}_{11}}{{\cal A}_{12}} \right |a_2 ,~~
a_3=\fr{1}{2r_{\nu 3}}\left |\fr{{\cal A}_{33}}{{\cal A}_{23}}\right |a_2,~~b_1=\fr{|{\cal A}_{12}|}{|\bar m|a_2}, ~~
b_3=\fr{\left |{\cal A}_{23} \right |}{r_{\nu 3}|\bar m|a_2}~.
\la{phi-yuk-AprB1}
\eeq
Using all these, we can calculate the baryon asymmetry. The results are given in Tab. \ref{nb-Aprb1IH}.
The goal of attaining needed baryon asymmetry with the minimal allowed value of $\tan\beta$ and without coming in contradiction with the
experimental data can be achieved in case of Normal Hierarchy as well by selecting:
\beq
\mathrm{For~ NH ~of ~the~{B_1}'~neutrino~ texture}:~~~ m_3=0.0741678 ~{\rm eV}~~~~ {\rm and}~~~~ \Delta\rho=3.2526  \la{m3delroB1NH}
\eeq
\begin{center}
  \begin{tabular}{|c|c|c|c|c|c|c|c|c|}
  \hline
 \multicolumn{1}{|c|}{\sffamily {\rm Case}}&\multicolumn{1}{|c|}{\sffamily {\rm M(GeV})}&\multicolumn{1}{|c|}{\sffamily $\tan\beta
 $}&\multicolumn{1}{|c|}{\sffamily $r_{\bar{m}}$}&\multicolumn{1}{|c|}{\sffamily
 $r_{v_{u}}$}&\multicolumn{1}{|c|}{\sffamily $\kappa_{N}$}&\multicolumn{1}{|c|}{\sffamily
 $10^{4}\times\xi_{\tau}$}&\multicolumn{1}{|c|}{\sffamily $ 10^{11}\!\times \!\l \!\frac{n_b^f}{s}\!\r_{\rm max} $}&\multicolumn{1}{|c|}{\sffamily $ 10^{11}\!\times \!\l \!\frac{n_b}{s}\!\r_{\rm max} $}\\
  \hline
\textbf{(I.1)}&$3\cdot10^3$ &2.1  &0.8928 &0.9731 &1.118  &0.8134 &8.57 & 8.62\\
  \hline
\textbf{(I.2)}&$10^4$ &2.135  &0.8499  &0.9574 &1.0986  &0.7826  & 8.55&8.6 \\
  \hline
\textbf{(I.3)}&$10^5$&2.332  &0.7856&0.9316  &1.0545   &0.7924   &8.56 & 8.61 \\
  \hline
\textbf{(I.4)}&$10^6$&2.559 &0.7385 &0.9103 &1.0209   &0.8066 & 8.56&8.6 \\
  \hline
\textbf{(I.5)}&$10^7$ &2.822&0.7048 &0.8926  &0.9959  &0.8242 & 8.54&8.59 \\
  \hline
  \hline
\textbf{(II.1)}&$6\cdot10^3$ &2.118  &0.875 &0.9695 &1.0933  &0.8109&8.55 &  8.6\\
  \hline
\textbf{(II.2)}&$10^4$ &2.119 &0.858  &0.9631 &1.0876 &0.7896 &8.56 &8.6 \\
  \hline
\textbf{(II.3)}&$10^5$&2.302  &0.7948&0.9378  &1.0481  &0.7932  &8.56 & 8.6 \\
  \hline
\textbf{(II.4)}&$10^6$&2.524 &0.7484 &0.9168&1.017 &0.8067 &8.55 &8.59 \\
  \hline
\textbf{(II.5)}&$10^7$ &2.786  &0.715 &0.8994  &0.9936  &0.826 &8.55 &8.59 \\
  \hline
  \end{tabular}
 \captionof{table}{${\rm B_1}'$ Neutrino Texture, IH. Baryon asymmetry for various values of $M$ and for corresponding minimal (allowed) values
 of $\tan\beta $.
With the choice given in Eqs. (\ref{m3delroB1IH}), (\ref{mi-phasesB1IH}) and bfv's of  $s^2_{ij}$. With
$\phi = -2.9846$ and for all cases $r_{\nu 3}\simeq 1$.}
 \label{nb-Aprb1IH}
\end{center}
give:
\beqs
\l m_1, m_2, m_3\r =(0.05437, ~0.0550533, ~0.0741678)~{\rm eV},
\eeqs
\beq
 ~~~\l \de , \rho_1, \rho_2 \r =\l 0.0034537,~ 0.25965,~ -2.9929\r .
\la{mi-phasesB1NH}
\eeq
\beq
\phi=2.2568  ,~~~~m_{\bt \bt}\simeq 0.051~{\rm eV}.\la{btphi}
\eeq
\begin{center}
  \begin{tabular}{|c|c|c|c|c|c|c|c|}
  \hline
 \multicolumn{1}{|c|}{\sffamily {\rm Case}}&\multicolumn{1}{|c|}{\sffamily {\rm M(GeV})}&\multicolumn{1}{|c|}{\sffamily $\tan\beta
 $}&\multicolumn{1}{|c|}{\sffamily $r_{\bar{m}}$}&\multicolumn{1}{|c|}{\sffamily
 $r_{v_{u}}$}&\multicolumn{1}{|c|}{\sffamily $\kappa_{N}$}&\multicolumn{1}{|c|}{\sffamily
 $10^{4}\times\xi_{\tau}$}&\multicolumn{1}{|c|}{\sffamily $ 10^{11}\!\times \!\l \!\frac{n_b}{s}\!\r_{\rm max} $}\\
  \hline
\textbf{(I.1)}&$3\cdot10^3$ &12.612  &0.9047 & 0.9764 &1.0026 &23.596&  8.6\\
  \hline
\textbf{(I.2)}&$10^4$ & 12.081  &0.8733 &0.9639 &0.9929  &20.327   &8.6  \\
  \hline
\textbf{(I.3)}&$10^5$ & 12.355  &0.8229 &0.9425 &0.9772  &18.774  & 8.6 \\
  \hline
\textbf{(I.4)}&$10^6$ & 12.696   &0.7829  &0.9236  &0.9652  &17.364 &8.6\\
  \hline
\textbf{(I.5)}&$10^7$ &13.066 &0.7515 &0.9071  &0.9566  &15.947  &8.6 \\
  \hline
   \hline
\textbf{(II.1)}&$6\cdot10^3$ & 12.608  &0.8858 &0.9725 &0.994 &23.269 &  8.6\\
  \hline
\textbf{(II.2)}&$10^4$ & 12.158  &0.8735  &0.9675 &0.9904  &21.059  &8.6  \\
  \hline
\textbf{(II.3)}&$10^5$ &12.249  &0.8253&0.9467  &0.9757   &18.883 & 8.6 \\
  \hline
\textbf{(II.4)}&$10^6$ &12.582  &0.787  &0.9284  &0.9645   &17.46  &8.6 \\
  \hline
\textbf{(II.5)}&$10^7$ & 12.943  &0.7567 &0.9122 &0.9565   &16.029 &8.6 \\
  \hline
  \end{tabular}
 \captionof{table}{${\rm B_1}'$ Neutrino Texture, NH. Baryon asymmetry for various values of $M$ and for corresponding minimal (allowed) values
 of $\tan \beta $.
With the choice given in Eqs. (\ref{m3delroB1NH}), (\ref{mi-phasesB1NH}) and bfv's of  $s^2_{ij}$. With $\phi = 2.2568$ and
for all cases $r_{\nu 3}\simeq 1$ and
$\frac{\tilde n_b}{s}\simeq 0$.}
 \label{nb-Aprb1NH}
\end{center}
The baryon asymmetries for cases corresponding to this NH scenario are given in Tab. \ref{nb-Aprb1NH}.

\vspace{0.3cm}

{\bf Texture B$_2$}

\vspace{0.3cm}

This texture is interesting because, due to specific form of
$Y_{\nu }$, the radiative corrections through the $\lam_{\tau }$ coupling do not generate cosmological CP asymmetry.
Thus $\lam_{\mu}$ may be important, which we investigate below. Thus, this model (and its slight modification discussed below) serves as a
good demonstration of the role of $\xi_{\mu }$ correction in emergence of needed Baryon asymmetry.

The B$_2$ Yukawa texture can be written as:
 \beq
{\rm Texture~~ B_2}:~~Y_{\nu }=\begin{pmatrix}
a_{1}e^{i\alpha_{1}} & b_{1}e^{i\beta_{1}}\\
a_{2}e^{i\alpha_{2}} & b_{2}e^{i\beta_{2}}\\
a_{3}e^{i\alpha_{3}}  &  0
\end{pmatrix}
=
\begin{pmatrix}
e^{ix} & 0&0\\
0 & e^{iy}&0\\
0 & 0&e^{iz}
\end{pmatrix}
\begin{pmatrix}
a_{1} & b_{1}\\
a_{2} &  b_{2}e^{i\phi}\\
a_3 &0
\end{pmatrix}
\begin{pmatrix}
e^{i\omega} & 0\\
0 & e^{i\rho}
\end{pmatrix}, \nonumber
\eeq
$$
{\rm with}   ~~~~x= \beta_{1}-\rho, \quad
y= \alpha_{2}-\alpha_{1}+\beta_{1}-\rho, \quad  z= \alpha_{3}-\alpha_{1}+\beta_{1}-\rho,
$$
\beq
\omega=\alpha_{1}-\beta_{1}+\rho, \quad \phi=\alpha_{1}-\beta_{1}-\alpha_{2}+\beta_{2}.
\la{YB2}
\eeq
Via the see-saw we will get the light neutrino mass matrix:
\beq
M_{\nu}^{(\rm B_2)}(M_Z)=\l \!\begin{array}{ccc}
2a_1b_1 &a_1b_2e^{i\phi }\!+\!a_2b_1& a_3b_1r_{\nu 3}\\
a_1b_2e^{i\phi }\!+\!a_2b_1 &2a_2b_2e^{i\phi }& a_3b_2e^{i\phi }r_{\nu 3}\\
a_3b_1r_{\nu 3}& a_3b_2e^{i\phi }r_{\nu 3} &0\end{array}\!\r \!\bar m~.
\la{Mnu-B2}
\eeq
This  neutrino mass texture (referred as B$_2$ neutrino texture) works only for inverted neutrino mass ordering \cite{Babu:2008kp} (with $m_3=0$) and
has two  predictive relations. In particular, in terms of measured oscillation parameters we can calculate
the phases $\de $ and $\rho_1$.
The exact expressions are:
$$
\cos \de =\fr{m_1t_{12}^2t_{23}^2-m_2(t_{23}^2+t_{12}^2s_{13}^2)}{2(m_1+m_2)t_{12}t_{23}s_{13}}~ ,~~~~~
\rho_1=\pi -{\rm Arg}\l \fr{t_{12}t_{23}-s_{13}e^{i\delta }}{t_{23}+t_{12}s_{13}e^{i\delta }}\r^{\!\!2} ,
$$
\beq
{\rm with}~~~~~m_1=\sqrt{\Delta m_{atm}^{2}-\Delta m_{sol}^{2}} ,~~~~~~m_2=\sqrt{\Delta m_{atm}^{2}} ,~~~~~~m_3=0.
\la{pred-B2}
\eeq
From these relations one can easily check that model works only if at least two of the oscillation parameters $\sin^2\te_{ij}$
are off by several  $\sigma$'s. Taking bfv's of the oscillation parameters would give the absolute values of
the r.h.s. of expression for $\cos \de $ larger than one. Besides this difficulty, proper value of the baryon asymmetry (generated with
help of 1-loop correction of $\lam_{\mu }$) requires even more deviation from the bfv's of the oscillation parameters.
The root of the problem is that the value of the phase $\phi $ is fixed so that the parameter $\sin \phi$ (governing cosmological CP
asymmetry) turns out to be too suppressed. For instance, with  $s_{12}^2=0.333$,
$s_{23}^2= 0.388$, $s_{13}^2= 0.0241$ and bfv's of
$\Delta m^2_{atm}$, $\Delta m^2_{sol}$, for $M=3\cdot 10^3$~GeV, with $\tan \beta\simeq 68$ and $M_S=1$~TeV we obtain needed baryon asymmetry
[$\l \fr{n_b}{s}\r_{max}\simeq 8.56\tm 10^{-11}$], however  for this case the values of  $\sin^2\te_{ij}$ are deviated
from the bfv's by $(2-3)\sigma $.

\vspace{0.3cm}

{\bf ${\rm B_2}'$ Neutrino Texture: Improved Version}

\vspace{0.3cm}
In order to avoid difficulties with B$_2$ neutrino texture  we add $d_5$ term to the  $(1,2)$ and $(2,1)$ elements of the light neutrino mass
matrix. After this, the $M_{\nu }$ will have the form:
\beq
M_{\nu}^{({\rm B_2}')}(M_Z)=\l \!\begin{array}{ccc}
2a_1b_1 &a_1b_2e^{i\phi }\!+\!a_2b_1\!\!+\!d_5& a_3b_1r_{\nu 3}\\
a_1b_2e^{i\phi }\!+\!a_2b_1\!\!+\!d_5&2a_2b_2e^{i\phi }& a_3b_2e^{i\phi }r_{\nu 3}\\
a_3b_1r_{\nu 3}& a_3b_2e^{i\phi }r_{\nu 3} &0\end{array}\!\r \!\bar m~.
\la{Mnu-B2pr}
\eeq
With this modification, all masses are non-zero, and therefore
 two additional parameters $m_3\neq 0$ and $\rho_2$ enter. We refer to this (\ref{Mnu-B2pr}) improved version as
the ${\rm B_2}'$  neutrino texture.
 Thus our relations will involve two more independent quantities.
For convenience we take $m_3$ and $\Delta \rho =\rho_1 -\rho_2 $ as such. From the condition $M_{\nu}^{(3,3)}=0$
we have:
\beq
m_1\left | U_{31}\right |^2\!=\!\left | m_2(U_{32})^2\!+\!m_3(U_{33})^2 \!e^{i\Delta \rho} \right |,~~
\rho_1\!=\!\pi \!-\!{\rm Arg}\lq \!\fr{m_2\left ( U_{31}\right )^2}{m_2(U_{32})^2\!+\!m_3(U_{33})^2 \!e^{i\Delta \rho}}\!\rq
~~~{\rm with}~~~\Delta \rho =\rho_1 -\rho_2.
\la{B2pr}
\eeq
 From these relations  the phases $\delta $ and $\rho_1$
can be calculated in terms of $m_3$ and $\Delta \rho $.

As it turns out, in this improved version the IH case works well for both neutrino sector and  the baryon asymmetry. So, we will
start with discussing the IH case.
For measured oscillation parameters we take the best fit values given in (\ref{ever-bfv}) and select pairs
$(m_3, \Delta \rho )$ in such a way as to get needed baryon asymmetry.
One such choice is:
\beq
m_3=0.01406~{\rm eV} ,~~~\Delta \rho =3.5257 ~,
\la{m3-delro-choice1}
\eeq
which with help of (\ref{ih1}) and (\ref{B2pr}) determine neutrino masses and phases  as:
\beqs
\l m_1, m_2, m_3\r =(0.0516, ~0.052323, ~0.01406)~{\rm eV},
\eeqs
\beq
 ~~~\l \de , \rho_1, \rho_2 \r =\l 2.8528,~ 3.1385,~ -0.38724\r .
\la{mi-phases}
\eeq
These for the observable $\nu 02\bt $-decay give $m_{\bt \bt}\simeq 0.0193$~eV.

As far as the baryon asymmetry is concerned, using (\ref{Mnu-B2pr}) in (\ref{nu1}) for the CP phase $\phi $ and expressing
couplings $a_{2,3}, b_{1,2}$ in terms of $a_1$
we get
$$
\phi ={\rm Arg}\l \fr{{\cal A}_{23}^2{\cal A}_{11}}{{\cal A}_{13}^2{\cal A}_{22}}\r ,
$$
\beq
a_2=\left |\fr{{\cal A}_{22}{\cal A}_{13}}{{\cal A}_{11}{\cal A}_{23}} \right |a_1 ,~~
a_3=\fr{2}{r_{\nu 3}}\left |\fr{{\cal A}_{13}}{{\cal A}_{11}}\right |a_1,~~b_1=\fr{|{\cal A}_{11}|}{2|\bar m|a_1}, ~~
b_2=\left |\fr{{\cal A}_{23}{\cal A}_{11}}{{\cal A}_{13}} \right |\fr{1}{2|\bar m|a_1}~.
\la{phi-yuk-B2pr}
\eeq
For the values of (\ref{m3-delro-choice1}), (\ref{mi-phases}) and bfv's for the $\te_{ij}$ angles we get
\beq
\phi =2.2301~.
\la{phi-B2pr-1}
\eeq
With these, and for given values of $M$ and $\tan \bt $ by varying $a_1$ we can investigate the baryon asymmetry.
Results are given in Tab. \ref{nb-B2pr}.

%
\begin{table}
\vs{0.3cm}
 $$\begin{array}{|c|c|c|c|c|c|c|c|c|c|}
\hline
{\rm Case}  & {\rm M(GeV}) & \tan \bt  &  r_{\nu 3} & r_{\bar m} & r_{v_u} & \ka_N &
 10^{4}\!\tm \! \xi_{\mu }  & 10^{11}\!\tm \!\l \!\fr{n_b^f}{s}\!\r_{\rm max} & 10^{11}\!\tm \!\l \!\fr{n_b}{s}\!\r_{\rm max} \\
\hline
 {\bf (I.1)} &  3\cdot 10^3 &69.256 &0.9965 &0.9048 &0.9763 & 1.047 &4.18 &8.55 & 8.6 \\
 \hline
 {\bf (I.2)} &  10^4 & 67.557 &0.9929 & 0.8728 &0.9638 &1.0327 &3.589 &8.55 & 8.6\\
\hline
 {\bf (I.3)} &  10^5 &67.376 & 0.9854 &0.8196 &0.9415  & 1.0176 &3.34 &8.55 & 8.6 \\
 \hline
 {\bf (I.4)} &  10^6 &67.359  & 0.9771 &0.7749& 0.9213 &1.006 & 3.122 &8.55 & 8.6 \\
 \hline
 {\bf (I.5)} &  10^7 &67.376&0.9681&0.7373& 0.9027 &0.997 &2.903&8.56 & 8.6 \\
 \hline
 \hline
 {\bf (II.1)} &  6\cdot 10^3 &70.391 & 0.9964 & 0.8858 & 0.9725 &1.0311 &4.093 &8.55 & 8.6 \\
 \hline
 {\bf (II.2)} &  10^4 & 69.003 &0.9949 &0.8735 & 0.9675 & 1.0243 & 3.691 & 8.55& 8.6\\
\hline
 {\bf (II.3)} &  10^5 & 68.322& 0.9873 & 0.8234 & 0.9462 & 1.0094& 3.33 &8.55 & 8.6 \\
 \hline
 {\bf (II.4)} &  10^6 &68.321  & 0.979 & 0.7813 & 0.9267 &0.9988  & 3.108&8.55 & 8.6 \\
 \hline
 {\bf (II.5)} &  10^7 & 68.373 & 0.9699 &0.7459 & 0.909 & 0.9907 & 2.889&8.56 & 8.61 \\
 \hline
\end{array}$$
\vs{-0.5cm}
\caption{${\rm B_2}'$ Neutrino Texture, IH neutrinos. Baryon asymmetry for various values of $M$ and for corresponding minimal (allowed) values
of $\tan \bt $.
For the values of (\ref{m3-delro-choice1}), (\ref{mi-phases}) and bfv's of $\te_{ij}$ mixing angles.}
 \vs{-0.3cm}
 \label{nb-B2pr}
\end{table}
%
%
%

As far as the NH case is concerned, the neutrino sector can work well by
certain selection of $(m_3, \Delta \rho )$. However, in order to generate needed baryon asymmetry we need to take values of $\sin^2\te_{ij}$
deviated from the bfv's by the $(2-3)\si $. For example, with
$(s^2_{12} , s^2_{23}, s^2_{13})=(0.27, 0.629, 0.022)$ and $(m_3, \Delta \rho )=(0.060651~{\rm eV}, 3.12)$
we get
$$
{\rm for}~{\rm NH} ~{\rm of~the}~{\rm B_2}'~{\rm neutrino ~texture}: \hspace{0.9cm}\l m_1, m_2, m_3\r =(0.033671,0.034764,0.060651)~{\rm eV},
$$
\beq
\l \de , \rho_1, \rho_2 \r =\l -0.013,~ -0.12393,~ 3.0393\r ~~
\Longrightarrow  ~~ \phi =-2.7538,~~ m_{\bt \bt }\simeq 0.032~{\rm eV}.
\la{Bpr-NH}
\eeq
These for $\tan \bt =68.1$ and $M=10^6$~GeV, $M_S=1$~TeV give the baryon asymmetry $\l \fr{n_b}{s}\!\r_{\rm max}\simeq 8.59\cdot 10^{-11}$.

Note that the ${\rm B_2}'$ neutrino texture  coincides with the texture $P_7$ of Ref. \cite{Achelashvili:2016nkr}  if all entries
 in (\ref{Mnu-B2pr}) are taken to be real.
As was shown in \cite{Achelashvili:2016nkr} the real neutrino mass texture with $M_{\nu}^{(3,3)}=0$ will work for both NH and IH neutrinos (see
Tab. 6 of Ref. \cite{Achelashvili:2016nkr}).
Advantage of complex $d=5$ entry [like in texture (\ref{Mnu-B2pr})] is that it gives good possibility for generation of the baryon asymmetry
with the $\lambda_{\mu }$'s radiative correction playing the decisive role. Similar possibility has not been considered in the
literature before.

Concluding, note also that the ${\rm A}'$ and ${\rm B_1}'$ neutrino textures  are generalizations of the textures $P_5$ and $P_6$ (respectively),
considered in \cite{Achelashvili:2016nkr}. The latter two had no complex phases, while ${\rm A}'$ and ${\rm B_1}'$ scenarios besides good neutrino
fits give possibility for the generation of the baryon asymmetry.
\section{Discussion and Outlook} \la{discussions}

In this work we have investigated the resonant leptogenesis within the extension of the MSSM by two right handed neutrino superfields
with quasi-degenerate masses ${\stackrel{<}{_\sim}10^7}$~GeV. It was shown that in this regime the cosmological CP asymmetry
arises at one loop level due to charged lepton Yukawa couplings. In particular, needed corrections may come from either of the $\lambda_{\tau}$ and $\lambda_{\mu}$ couplings. Which one is relevant from these two couplings depends on the structure of the $3\times 2$ Dirac type Yukawa
 matrix $Y_{\nu }$.
 Aiming to make close connection with the neutrino sector, we first examined all viable neutrino models (considered earlier in Ref. \cite{Achelashvili:2016nkr})
  based on two texture zero $Y_{\nu }$'s augmented by single $\Delta L=2$, ${\rm d}=5$ operators. This setup is predictive and allows
  to relate leptonic CP violating phase $\delta $ with the cosmological CP violation. In one of such scenarios the role of
  the $\lambda_{\mu }$ coupling in CP asymmetry generated at quantum level has been demonstrated. We have also revised the
  models of Ref. \cite{Babu:2008kp} and considered their improved versions by including proper $\Delta L=2$, ${\rm d}=5$ operators.
 This allowed to have good fit with the neutrino data and generate needed amount of the baryon asymmetry.

 Without specifying their origin, in our considerations we have extensively applied the $\Delta L=2$, ${\rm d}=5$ operators, of the form given in Eq. (\ref{d5}). Such $\rm d=5$ couplings can be generated from a different sector via renormalizable interactions.
For instance, introducing the pair of MSSM singlet states ${\cal N}$, $\overline{\cal N}$ and the superpotential couplings
\begin{equation}
\lam^{(i)}l_i{\cal N}h_u+\bar \lam^{(j)}l_j\overline{\cal N}h_u-M_*{\cal N}\overline{\cal N}~,
\end{equation}
it is easy to verify that integration of the heavy ${\cal N}$, $\overline{\cal N}$ multiplets leads to the operator in Eq. (\ref{d5})
with
\begin{equation}
\tilde {d_5}e^{ix_5}=2\lam^{(i)}\bar \lam^{(j)}~.
\end{equation}
Important ingredient here is to maintain forms of the  matrices $Y_{\nu }$, $M_N$. In \cite{Achelashvili:2016trx} considering one such fully consistent extension, it was demonstrated that all obtained results (e.g. neutrino masses and mixings, and baryon asymmetry as well) can remain intact.
Although the way demonstrated above is rather simple, there can be considered also alternative ways for
generating those $\Delta L=2$ effective couplings.
These could be done either in a spirit of type II \cite{Magg:1980ut}, or  type III   \cite{Foot:1988aq} see-saw
mechanisms, or even exploiting alternative possibilities \cite{alt-radiative}, \cite{alt-high-d-ops} through the introduction
of appropriate extra states. Details of such scenarios should be pursued elsewhere.

Throughout our studies we have studied texture zero coupling matrices, but did not
 attempt to explain and justify  considered structures by symmetries. Our approach, being rather phenomenological, was to consider such textures which give predictive and/or consistent scenarios allowing for transparent demonstrations of the suggested mechanism of the
 loop induced cosmological CP violation.
 It is desirable to have explanation of texture zeros at more fundamental level, and exploiting flavor symmetries
 seems to be a good framework. We are planning to pursue this approach  in a future work \cite{our-in-prep}.

Since the supersymmetry is a well motivated construction, we have  performed our investigations within its framework.
However, it would be interesting to examine the considered models also within the non-SUSY setup. For the latter, the scenarios with
 low $\tan \bt$ look encouraging to start with.

Finally, it would be challenging to embed considered models in Grand Unification (GUT) such as $SU(5)$ and $SO(10)$ GUTs.
Due to the high GUT symmetries, additional relations and constraints would emerge making models more predictive.
 These and related issues will be addressed elsewhere.

\vspace{0.2cm}
{\it Note added:} After this paper was submitted to arXiv, we have been informed by F.R. Joaquim  that the
role of $\lambda_{\tau }$ coupling for the resonant leptogenesis within non-SUSY scenarios had been investigated
in earlier works \cite{early-tau-effect}.

\subsubsection*{Acknowledgments}

Z.T. thanks CERN theory division for warm hospitality and partial support during his visit there.

\appendix

\renewcommand{\theequation}{A.\arabic{equation}}\setcounter{equation}{0}

\section{Renormalization Group Studies}
\la{app-RG}

\subsection{Running of $Y_{\nu }, Y_e$ and $M_N$ Matrices}
\la{app-YM-RGs}

RG equations for the charged lepton and neutrino Dirac Yukawa matrices, appearing in the superpotential of Eq. (\ref{r21}), at 1-loop order
have the forms
\cite{Martin:1993zk}, \cite{Antusch:2002ek}:
\beq
16\pi^2 \fr{d}{dt}Y_e=3Y_eY_e^\dag Y_e+Y_{\nu}Y_{\nu}^\dag Y_e+
Y_e \left [ {\rm tr}\l 3Y_d^\dag Y_d+Y_e^\dag Y_e \r
 -c_e^ag_a^2\right ]~,~~~~~c_e^a=(\fr{9}{5}, 3, 0) ,
\la{Ye-RG}
\eeq
\beq
16\pi^2 \fr{d}{dt}Y_{\nu}=Y_eY_e^\dag Y_{\nu}+3Y_{\nu}Y_{\nu}^\dag Y_{\nu}+
Y_{\nu} \left [ {\rm tr}\l 3Y_u^\dag Y_u+Y_{\nu}^\dag Y_{\nu} \r
 -c_{\nu }^ag_a^2\right ]~,~~~~~c_{\nu }^a=(\fr{3}{5}, 3, 0) .
\la{Ynu-RG}
\eeq
$g_a=\l g_1, g_2, g_3\r $ denote gauge couplings of $U(1)_Y, SU(2)_w$ and $SU(3)_c$ gauge groups respectively. Their 1-loop RG have forms
$16\pi^2 \fr{d}{dt}g_a=b_ag_a^3$, with $b_a=(\fr{33}{5}, 1, -3)$, where the hypercharge of $U(1)_Y$ is taken in $SU(5)$ normalization.

The RG for the RHN mass matrix at 2-loop level has the form \cite{Antusch:2002ek}:
$$
16\pi^2\fr{d}{dt}M_N=2M_NY_{\nu}^{\dag}Y_{\nu}
-\fr{1}{8\pi^2}M_N\left [ Y_{\nu}^{\dag}Y_eY_e^{\dag}Y_{\nu}+Y_{\nu}^{\dag}Y_{\nu}Y_{\nu}^{\dag}Y_{\nu}+
Y_{\nu}^{\dag}Y_{\nu}{\tr}(3Y_u^\dag Y_u+Y_{\nu}^{\dag}Y_{\nu})\right ]
$$
\beq
+\fr{1}{8\pi^2}M_NY_{\nu}^{\dag}Y_{\nu}\l \fr{3}{5}g_1^2+3g_2^2\r +({\rm transpose})~,
\la{MN-2loop-RG}
\eeq

Let's start with  renormalization of the $Y_{\nu }$'s matrix elements.
Ignoring in Eq. (\ref{Ynu-RG}) the ${\cal O}(Y_{\nu}^3)$ order entries (which are very small because within our studies
$|(Y_{\nu})_{ij}|\stackrel{<}{_\sim }10^{-4}$), and from charged fermion Yukawas keeping $\lam_{\tau}$, $\lambda_{\mu }$, $\lam_t$ and $\lam_b$, we
will have:
\beq
16\pi^2 \fr{d}{dt}\ln (Y_{\nu})_{ij}\simeq \de_{i3}\lam_{\tau }^2+\de_{i2}\lam_{\mu }^2+3\lam_t^2-c_{\nu }^ag_a^2~.
\la{Ynu-RG-approx}
\eeq
This gives the solution
\beq
(Y_{\nu})_{ij}(\mu)=(Y_{\nu G})_{ij}(\eta_{\tau}(\mu))^{\de_{i3}}(\eta_{\mu}(\mu))^{\de_{i2}}\eta_t^3(\mu)\eta_{g\nu }(\mu ) ,
\la{approx-Ynu-sol}
\eeq
where $Y_{\nu G}$ denotes Yukawa matrix at scale $M_G$ and the scale dependent RG factors are given by:
$$
\eta_{t, b, \tau, \mu}(\mu)\!=\!\exp \!\l \!\!-\fr{1}{16\pi^2}\!\!\int_t^{t_G}\!\!\lam^2_{t, b, \tau, \mu } (t')dt' \!\r ,~~
\eta_a(\mu )\!=\!\exp \!\l \!\!\fr{1}{16\pi^2}\!\!\int_t^{t_G}\!\!\! g^2_a (t')dt' \!\r
$$
\beq
\eta_{g\nu }(\mu )\!=\exp \l \!\!\fr{1}{16\pi^2}\!\int_t^{t_G}\!\!\!\!c_{\nu}^ag^2_a (t')dt' \!\!\r
=\eta_1^{3/5} (\mu ) \eta_2^3 (\mu ),~~~
{\rm with}~~~t=\ln \mu ~,~t'=\ln \mu' ~,~~t_G=\ln M_G .
\la{RG-factors}
\eeq
From these, for the combination $Y_{\nu }^\dag Y_{\nu }$ at scale $\mu =M$ we get expression given in Eq. (\ref{YYnu-M}).

On the other hand, for the RHN mass splitting and for the phase mismatch [depending on $\xi_{\tau, \mu }$ defined in Eq. (\ref{xi-shift})], the
integrals/factors of Eqs. (\ref{int-matrix}), (\ref{bar-r-kapa-2loop}),  (\ref{r-kapa})
and (\ref{YYnu-M}) will be relevant.

\subsection{Relating $M_{\nu }(M_Z)$ and $M_{\nu }(M)$}
\la{app-nuRG}
Details of derivations, of the results presented in this subsection, are given in Appendix A.2 of Ref. \cite{Achelashvili:2016trx}. At scale $M$, after decoupling of the RHN states, the neutrino mass matrix is generated and has the form:
\beq
M_{\nu }^{ij}(M)=-
\left(
  \begin{array}{ccc}
    \tm & \tm & \tm \\
    \tm & \tm & \tm \\
    \tm & \tm & \tm \\
  \end{array}
\right)\fr{v_u^2(M)}{M e^{-i(\omega +\rho )}} ~,
\la{Mnu-M}
\eeq
where `$\times $' stand for entries depending on Yukawa couplings. After renormalization, keeping $\lam_{\tau}, \lam_t$, $\lambda_{b}$ and $g_a$ in the RGs,
 the neutrino mass matrix at scale $M_Z$ has the form:
\beq
M_{\nu }^{ij}(M_Z)=
\left(
  \begin{array}{ccc}
    \tm & \tm & (\tm )\!\cdot \!r_{\nu 3}\\
    \tm & \tm & (\tm )\!\cdot \! r_{\nu 3}\\
    (\tm )\!\cdot \! r_{\nu 3}& (\tm )\!\cdot \! r_{\nu 3} & (\tm )\!\cdot \! r_{\nu 3}^2\\
  \end{array}
\right)\bar m ~,
\la{Mnu-MZ}
\eeq
with $\bar m$ given in Eq. (\ref{bar-m-rbar}) and $\times $ in Eq. (\ref{Mnu-MZ}) denotes entries determined at scale $M$ and corresponding to those in (\ref{Mnu-M}), and
RG factors $r_{\nu 3}$, $r_{\bar m}$ are given respectively in Eqs. (A.17), (A.18) of Ref. \cite{Achelashvili:2016trx}.

We will also need the RG factor relating the VEV $v_u(M)$ to  the $v(M_Z)$. Thus we define:
\beq
r_{v_u}\!=\!\fr{v_u(M)}{v(M_Z)s_{\bt }}\! ~.
\la{r-vu}
\eeq
Analytic expression for $r_{v_u}$ derived from appropriate RGs is given by Eq. (A.20) of Ref. \cite{Achelashvili:2016trx}.

\subsection{Calculation Procedure and Used Schemes}
\la{app-baund-match}

To find the RG factors, appearing in the baryon asymmetry and in the neutrino mass matrix renormalization, we  numerically solve renormalization group equations from
the scale $M_Z$ up to the $M_G\simeq 2\cdot 10^{16}$~GeV scale. For simplicity, for all SUSY particle masses we take common mass scale $M_S$.
Thus, in the energy interval $M_Z\leq \mu <M_S$, the Standard Model RGs for  $\ov{\rm MS}$ coupling constants are used.
However, in the interval  $M_S\leq \mu \leq M_G$, since we are dealing with the SUSY, the RGs for the $\ov{\rm DR}$ couplings are applied.
Below we give boundary and matching conditions for the gauge couplings $g_{1,2,3}$, for Yukawa constants $\lam_{t,b,\tau,\mu}$ and for the Higgs
self-coupling $\lam $.

\vs{0.2cm}
\hs{-0.5cm}{\bf Gauge couplings $\alpha_a=\frac{g_a^2}{4\pi }$}

We choose our inputs for the $\ov{\rm MS}$ gauge couplings at scale $M_Z$ as follows:
$$
\al_1^{-1}(M_Z)=\fr{3}{5}c_w^2\al_{em}^{-1}(M_Z)+\fr{3}{5}c_w^2\fr{8}{9\pi }\ln \fr{m_t}{M_Z}~,~~~~~
\al_2^{-1}(M_Z)=s_w^2\al_{em}^{-1}(M_Z)+s_w^2\fr{8}{9\pi }\ln \fr{m_t}{M_Z}~,
$$
\beq
\al_3^{-1}(M_Z)=\al_s^{-1}(M_Z)+\fr{1}{3\pi }\ln \fr{m_t}{M_Z}~,
\la{in-alpMZ}
\eeq
where logarithmic terms $\ln \fr{m_t}{M_Z}$ are due to the top quark threshold correction \cite{Hall:1980kf}, \cite{Arason:1991ic}.
Taking $\al_s(M_Z)=0.1185$, $\al_{em}^{-1}(M_Z)=127.934$ and $s_w^2=0.2313$, from (\ref{in-alpMZ}) we obtain:
$$
\al_1^{-1}(M_Z)=59.0057+\fr{8c_w^2}{15\pi }\ln \fr{m_t}{M_Z}~,~~~
\al_2^{-1}(M_Z)=29.5911+\fr{8s_w^2}{9\pi }\ln \fr{m_t}{M_Z}~,
$$
\beq
\al_3^{-1}(M_Z)=8.4388+\fr{1}{3\pi }\ln \fr{m_t}{M_Z}~.
\la{num-in-alpMZ}
\eeq
With these inputs we run $g_{1,2,3}$ via the 2-loop RGs from $M_Z$ up to the scale $M_S$.

At scale $\mu =M_S$ we use the matching conditions between $\ov{\rm DR}-\ov{\rm MS}$ gauge couplings
\cite{{Antoniadis:1982vr},{Martin:1993yx}}:
\beq
{\rm at}~\mu =M_S:~~~
\fr{1}{\al_1^{\rm \ov{DR}}} = \fr{1}{\al_1^{\rm \ov{MS}}}~,~~~~\fr{1}{\al_2^{\rm \ov{DR}}} = \fr{1}{\al_2^{\rm \ov{MS}}}-\fr{1}{6\pi }~,~~~~
\fr{1}{\al_3^{\rm \ov{DR}}} = \fr{1}{\al_3^{\rm \ov{MS}}}-\fr{1}{4\pi }~.
\la{SUSY-alpha-DR-MS}
\eeq
Above the scale
$M_S$ we apply 2-loop SUSY RG equations in  $\ov{\rm DR}$ scheme \cite{Martin:1993zk}.

\vs{0.2cm}
\hs{-0.5cm}{\bf Yukawa Couplings and $\lam $}

At the scale $M_S$ all SUSY states decouple and we are left with the Standard Model with one Higgs doublet.
Thus, Yukawa couplings we are considering and the self-coupling are determined as:
$$
\lam_t(m_t)=\fr{m_t(m_t)}{v(m_t)}~,~~~~\lam_b(M_Z)=\fr{2.89{\rm GeV}}{v(M_Z)}~,~~~~\lam_{\tau}(M_Z)=\fr{1.746{\rm GeV}}{v(M_Z)}~,~ \lambda_{\mu }(M_Z)=
\frac{0.1027 {\rm GeV}}{v(M_Z)},~~~
$$
\beq
\lam(m_h)=\fr{1}{4}\l \fr{m_h}{v(m_h)}\r^2 ,~~~~~{\rm with}~~~v(M_Z)=174.1~{\rm GeV}~,~~~m_h=125.15~{\rm GeV}~,
\la{yuk-MZ-mt}
\eeq
where $m_t(m_t)$ is the top quark running mass related to the pole mass as:
\beq
m_t(m_t)=p_tM_t^{pole}~.
\la{run-pole-mt}
\eeq
The factor $p_t$ is $p_t\simeq 1/1.0603$ \cite{pole-run}, while the recent measured value of the top's pole mass is \cite{ATLAS:2014wva}:
\beq
M_t^{pole}=(173.34\pm 0.76)~{\rm GeV}.
\la{exp-top-pole}
\eeq
We take the values of (\ref{yuk-MZ-mt}) as boundary conditions for solving 2-loop RG equations \cite{Machacek:1983fi}, \cite{Arason:1991ic}
for $\lam_{t, b, \tau, \mu}$ and $\lam $ from the $M_Z$ scale up to the scale $M_S$.

Above the $M_S$ scale, we have MSSM states including two doublets
$h_u$ and $h_d$, which couple with up type quarks and down type quarks/charged leptons respectively.
Thus, Yukawa couplings we are considering at $M_S$ are $\approx \lam_t(M_S)/s_{\bt } ,  \lam_b(M_S)/c_{\bt }$
and $\lam_{\tau, \mu }(M_S)/c_{\bt }$, with $s_{\bt }\equiv \sin \bt, c_{\bt }\equiv \cos \bt $. Above the scale
$M_S$ we apply 2-loop SUSY RG equations in  $\ov{\rm DR}$ scheme \cite{Martin:1993zk}. Thus, at $\mu =M_S$ we use the matching conditions
between $\ov{\rm DR}-\ov{\rm MS}$ couplings:
$$
{\rm at}~\mu =M_S:~~~~~~~~~~~~~~\lam_t^{\rm \ov{DR}}\simeq \fr{\lam_t^{\rm \ov{MS}}}{s_{\bt }}
\left [ 1+\fr{1}{16\pi^2}\l \fr{g_1^2}{120}+\fr{3g_2^2}{8}-\fr{4g_3^2}{3}\r \right ] ,
$$
\beq
\lam_b^{\rm \ov{DR}}\simeq \fr{\lam_b^{\rm \ov{MS}}}{c_{\bt }}
\left [1\!+\!\fr{1}{16\pi^2}\l \fr{13g_1^2}{120}+\fr{3g_2^2}{8}-\fr{4g_3^2}{3}\r \right ],~~~~
\lam_{\tau , \mu }^{\rm \ov{DR}}\simeq \fr{\lam_{\tau , \mu }^{\rm \ov{MS}}}{c_{\bt }}
\left [1\!+\!\fr{1}{16\pi^2}\l -\fr{9g_1^2}{40}+\fr{3g_2^2}{8}\r \right ] ,
\la{SUSYyuk-DR-MS}
\eeq
where expressions in brackets of r.h.s. of the relations are due to the $\ov{\rm DR}-\ov{\rm MS}$ conversions \cite{Martin:1993yx}.
 With Eq. (\ref{SUSYyuk-DR-MS})'s matchings  we run corresponding couplings from the scale $M_S$ up to the $M_G$ scale.
 Throughout  the paper, above the mass scale $M_S$ without using
  the superscript $\ov{\rm DR}$ we assume the couplings determined in this scheme.
\renewcommand{\theequation}{B.\arabic{equation}}\setcounter{equation}{0}
\section{Baryon Asymmetry from RHS Decays}
\la{app-scalar-asym}

In this appendix we give details of the contribution to the net baryon asymmetry from the right handed sneutrinos (RHS) - the scalar partners
of the RHNs.
Estimation of this contribution for specific textures was given in \cite{Babu:2008kp}, while more detailed investigation was given in
\cite{Achelashvili:2016trx} (from the lepton couplings taking into account only $\lam_{\tau }$ and $A_{\tau }$ in the proper RGs).
Since we have seen that for some cases for the cosmological CP asymmetry decisive is the RG correction via the $\lam_{\mu }$ Yukawa coupling,
here we extend its calculation by taking into account also effects from $\lam_{\mu}$ and $A_{\mu }$ into the asymmetry generated by the RHS
decays.

We will consider soft SUSY breaking scalar potential
\beq
V_{SB}^{\nu}=\tl l^TA_{\nu}\tl Nh_u-\fr{1}{2}\tl N^TB_N\tl N+{\rm h.c.}+\tl l^{\dag}m^2_{\tl l}\tl l+\tl N^{\dag}m^2_{\tl N}\tl N~,
\la{V-SB-nu}
\eeq
which will be relevant for deriving RHS masses and their couplings to the components of the $l$ and $h_u$ superfields.
Using general expressions of Ref. \cite{Martin:1993zk} we write down 1-loop RGs for $A_{\nu}$ and $B_N$, which  have the forms:
$$
16\pi^2\fr{d}{dt}A_{\nu }=Y_eY_e^\dag A_{\nu }+2\hat A_eY_e^\dag Y_{\nu }+5Y_{\nu }Y_{\nu }^\dag A_{\nu }\!+\!
A_{\nu }\!\left [{\rm tr}(3Y_u^\dag Y_u+Y_{\nu }^\dag Y_{\nu })+4Y_{\nu }^\dag Y_{\nu }-c_{\nu}^ag_a^2\right ]
$$
\beq
+2Y_{\nu }\!\left [{\rm tr}(3Y_u^\dag \hat A_u+Y_{\nu }^\dag A_{\nu })+c_{\nu}^ag_a^2M_{\tl V_a}\right ]~,
\la{RG-Anu}
\eeq
\beq
16\pi^2\fr{d}{dt}B_N=2B_NY_{\nu }^\dag Y_{\nu }+2Y_{\nu }^T Y_{\nu }^*B_N+4M_NY_{\nu }^\dag A_{\nu }+4A_{\nu }^T Y_{\nu }^*M_N~.
\la{RG-B-N}
\eeq
We parameterize the matrices $B_N$ and $A_{\nu }$ as:
\beq
B_N=(M_N)_{12}m_B\left(
  \begin{array}{cc}
    \de_{BN}^{(1)} & 1 \\
    1 & \de_{BN}^{(2)} \\
  \end{array}
\right),~~~~~~~A_{\nu }=m_Aa_{\nu }~,
\la{BN-matrix}
\eeq
where  entries $(M_N)_{12}, m_B$, $\de_{BN}^{(1,2)}$ and elements of the matrix $a_{\nu }$ run (their RGs can be derived from the RG
equations given above), while $m_A$ is a constant.
The matrix $\hat A_e$ (similar to the structure of $Y_e$ Yukawa matrix) is
\beq
\hat A_e={\rm Diag}\l A_e, A_{\mu }, A_{\tau }\r .
\la{Ae-matr}
\eeq
Assuming proportionality / alignment of the soft SUSY breaking terms and corresponding superpotential couplings,
we will use the following boundary conditions:
$$
{\rm at}~\mu=M_G:~~~~a_{\nu }=Y_{\nu}~,~~~~\de_{BN}^{(1)}=\de_{BN}^{(2)}=0 ,~~~\hat A_e=m_A{\rm Diag }\l \lam_e, \lam_{\mu },\lam_{\tau }\r
$$
\beq
\hat A_u=m_AY_{uG}~,~~~~\hat A_d=m_AY_{dG}~.
\la{b-conds}
\eeq
Using  (\ref{RG-B-N}) for $B_N$'s entries in (\ref{BN-matrix}) we have:
\beq
16\pi^2\fr{d}{dt}\de_{BN}^{(1)}\simeq 4(Y_{\nu }^\dag Y_{\nu })_{21}+8\fr{m_A}{m_B}(Y_{\nu }^\dag a_{\nu })_{21} ~,~~~~
16\pi^2\fr{d}{dt}\de_{BN}^{(2)}\simeq 4(Y_{\nu }^\dag Y_{\nu })_{12}+8\fr{m_A}{m_B}(Y_{\nu }^\dag a_{\nu })_{12}.
\la{RG-de-BN}
\eeq
For the elements of $a_{\nu }$ we have
\beq
16\pi^2\fr{d}{dt}\l \fr{(a_{\nu })_{ij}}{(Y_{\nu })_{ij}}\r \simeq 2\fr{1}{m_A}(\de_{i3}\lam_{\tau}A_{\tau}+\de_{i2}\lam_{\mu }A_{\mu })+
\fr{2}{m_A}(3\lam_tA_t+c_{\nu }^ag_a^2M_{\tl V_a})~,
\la{RG-a-Y}
\eeq
which show the violation of the alignment between $a_{\nu }$ and $Y_{\nu }$ due to RG effects.
At r.h.s. of (\ref{RG-a-Y}) we kept $\lam_{\mu , \tau , t}, A_{\mu , \tau , t}$,  gauge couplings and gaugino masses.
From this we derive
$$
a_{\nu }\simeq
\left(
  \begin{array}{ccc}
    1+\ep_0 & 0 & 0 \\
    0 & 1+\ep_0+\ep_{\mu } & 0 \\
    0 & 0 & 1+\ep_0+\ep_{\tau } \\
  \end{array}
\right)\!Y_{\nu }
$$
\beq
{\rm with}~~~\ep_0= -\fr{1}{8\pi^2m_A}\int_{t}^{t_G} \!\!\!\!dt(3\lam_tA_t+c_{\nu }^ag_a^2M_{\tl V_a})~,~~~~~~
\ep_{\mu , \tau } =-\fr{1}{8\pi^2m_A}\int_{t}^{t_G} \!\!\!\!dt\lam_{\mu, \tau}A_{\mu, \tau }~.
\la{sol-a-nu}
\eeq
Keeping in mind that the powers of the $Y_{\nu }$ couplings can be ignored due to their smallness, the $m_B$ can be treated as a
constant, and from
(\ref{sol-a-nu}), (\ref{RG-de-BN}), (\ref{BN-matrix}) we obtain:
\beq
{\rm at}~\mu=M:~~~~B_N=m_BM\left(
  \begin{array}{cc}
    -\al \de_{N}(1+\bar{\ep }_1) & 1 \\
    1 & -\al \de_{N}^*(1+\bar{\ep }_2) \\
  \end{array}
\right),~~~~\al =1+2\fr{m_A}{m_B}~
\la{BN-at-M}
\eeq
and
$$
\bar{\ep }_1\!=\!\fr{1}{4\pi^2\al \de_{N}}\!\int_{t_M}^{t_G} \!\!dt\!\l \!\!Y_{\nu }^\dag (\fr{\al }{16\pi^2}Y_eY_e^\dag
+2\fr{m_A}{m_B}\hat{\ep })Y_{\nu }\!\r_{\!\!21}\!,~~~
\bar{\ep }_2\!=\!\fr{1}{4\pi^2\al \de_{N}^*}\!\int_{t_M}^{t_G} \!\!dt\!\l \!\!Y_{\nu }^\dag (\fr{\al^* }{16\pi^2}Y_eY_e^\dag
+2\fr{m_A^*}{m_B^*}\hat{\ep }^*)Y_{\nu }\!\r_{\!\!21}^{\!\!*}~,~
$$
\beq
{\rm with}~~~~\hat{\ep }={\rm Diag}\l \ep_0,~ \ep_0+\ep_{\mu },~ \ep_0+\ep_{\tau }\r .
\la{eps-12}
\eeq
The form of $B_N$ given in Eq. (\ref{BN-at-M}) will be used to construct the RHS mass matrix.  Before doing this, using Eq. (\ref{approx-Ynu-sol}) and ignoring the coupling $\lam_e$ (as it turns out from the lepton Yukawa couplings all relevant effects are due to $\lam_{\mu , \tau}$), for $\bar{\ep }_{1,2}$ at scale $\mu =M$ we can get expressions:

$$
\bar{\ep }_1(M)=\left. \fr{1}{4\pi^2\al \de_N}(Y_{\nu }^\dag \hat K Y_{\nu })_{21}\right |_{\mu =M},~~~~
\bar{\ep }_2(M)=\left. \fr{1}{4\pi^2\al \de_N^*}(Y_{\nu }^T \hat K Y_{\nu }^*)_{21}\right |_{\mu =M}
$$
$$
{\rm with}~~~~\hat K=\fr{1}{\eta_t^6\eta_{g\nu }^2}{\rm Diag}\!\left [ 2\fr{m_A}{m_B}I_0 ~,~
\fr{1}{\eta_{\mu }^2}\l 2\fr{m_A}{m_B}I_1^{(\mu )}+\fr{\al }{16\pi^2}I_2^{(\mu )} \r ~, ~
\fr{1}{\eta_{\tau }^2}\l 2\fr{m_A}{m_B}I_1^{(\tau )}+\fr{\al }{16\pi^2}I_2^{(\tau )} \r \right ],
$$
\beq
I_0=\int_{t_M}^{t_G}\!\!\!dt\eta_t^6\eta_{g\nu }^2\ep_0~,~~~~
I_1^{(\mu, \tau )}=\int_{t_M}^{t_G}\!\!\!dt\eta_t^6\eta_{g\nu }^2(\ep_0+\ep_{\mu, \tau })\eta^2_{\mu, \tau }~,~~~~
I_2^{(\mu, \tau )}=\int_{t_M}^{t_G}\!\!\!dt\eta_t^6\eta_{g\nu }^2\lambda^2_{\mu, \tau }\eta^2_{\mu, \tau }.
\la{ep12-K}
\eeq

Keeping the $B_N$-term in (\ref{V-SB-nu}) and including the mass$^2$ term $\tl N^\dag M_N^\dag M_N\tl N$ coming from the superpotential,
the quadratic (with respect to $\tl N$'s) potential will be:
\beq
V_{\tl N}^{(2)}=\tl N^\dag M_N^\dag M_N\tl N-\l \fr{1}{2}\tl N^TB_N\tl N+{\rm h.c. }\r ~.
\la{V-tlN-1}
\eeq
With the transformation of the $N$ superfields $N=U_NN'$ (according to Eq. (\ref{MN-diag-tion}), the $U_N$ diagonalizes the fermionic RHN
mass matrix), we obtain:
\beq
V_{\tl N}^{(2)}=\tl N^{\hs{0.5mm}'\!\dag } (M_N^{Diag})^2\tl N'-\l \fr{1}{2}\tl N^{\hs{0.5mm}'T }U_N^TB_NU_N\tl N'+{\rm h.c. }\r .
\la{V-tlN-2}
\eeq
With  phase redefinition
\beq
\tl N'=\tl P_1\tl N''~,~~~~~\tl P_1={\rm Diag}\l e^{-i\tl{\om}_1/2}, e^{-i\tl{\om}_2 /2}\r~,~~~~~{\rm with}~~~\tl{\om}_{1,2}={\rm
Arg}[m_B(1\mp \tl{\al}|\de_N|)]
\la{scN-phase}
\eeq
and by going to the real scalar components
\beq
{\tl{N}_1}''=\fr{1}{\sq{2}}(\tl{N}_1^R+i\tl{N}_1^I)~,~~~~~{\tl{N}_2}''=\fr{1}{\sq{2}}(\tl{N}_2^R+i\tl{N}_2^I)~,
\la{tlN-real}
\eeq
and using (\ref{BN-at-M}), we will have:
$$
-\l \fr{1}{2}\tl N^{\hs{0.5mm}'T }U_N^TB_NU_N\tl N'+{\rm h.c. }\r =-\fr{|Mm_B|}{2} \left |1-\tl{\al}|\de_N|\right |\l
\!(\tl{N}_1^R)^2-(\tl{N}_1^I)^2\!\r
$$
$$
-\fr{|Mm_B|}{2} \left |1+\tl{\al}|\de_N|\right |\l \!(\tl{N}_2^R)^2-(\tl{N}_2^I)^2\!\r
-|M|{\rm Re}(m_B\de_{\ep})\l \tl{N}_1^R\tl{N}_2^R-\tl{N}_1^I\tl{N}_2^I\r
+|M|{\rm Im}(m_B\de_{\ep})\l \tl{N}_1^I\tl{N}_2^R+\tl{N}_1^R\tl{N}_2^I\r
$$
\beq
{\rm with}~~~~\tl{\al}=\al (1+\fr{\bar{\ep }_1+\bar{\ep }_2}{2}),~~~  \de_{\ep}=i\al |\de_N|\fr{\bar{\ep }_1-\bar{\ep
}_2}{2}e^{-i(\tl{\om}_1+\tl{\om}_2)/2}~.
\la{BN-2}
\eeq
From (\ref{V-tlN-2}) and (\ref{BN-2}) we obtain the mass$^2$ terms:
\beq
V_{\tl N}^{(2)}=\fr{1}{2}\tl{n}^{0T}M_{\tl n}^2\tl n^0~,~~~~{\rm with}~~~~\tl{n}^{0T}=\l \tl{N}_1^R, \tl{N}_1^I, \tl{N}_2^R, \tl{N}_2^I\r ~
\la{V-tlN-3}
\eeq
and
\beq
M_{\tl n}^2=
\left(
  \begin{array}{cccc}
    (\tl M_1^0)^2 & 0 & -|M|{\rm Re}(m_B\de_{\ep}) & |M|{\rm Im}(m_B\de_{\ep}) \\
    0 & (\tl M_2^0)^2 & |M|{\rm Im}(m_B\de_{\ep}) & |M|{\rm Re}(m_B\de_{\ep}) \\
    -|M|{\rm Re}(m_B\de_{\ep}) & |M|{\rm Im}(m_B\de_{\ep}) & (\tl M_3^0)^2 & 0 \\
    |M|{\rm Im}(m_B\de_{\ep}) & |M|{\rm Re}(m_B\de_{\ep}) & 0 & (\tl M_4^0)^2 \\
  \end{array}
\right)
\la{Mtln}
\eeq
where
$$
(\tl M_1^0)^2=|M|^2(1-|\de_N|)^2-|m_BM|\left |1-\tl{\al}|\de_N|\right |,~~(\tl M_2^0)^2=|M|^2(1-|\de_N|)^2+|m_BM|\left
|1-\tl{\al}|\de_N|\right |~,
$$
\beq
(\tl M_3^0)^2=|M|^2(1+|\de_N|)^2-|m_BM|\left |1+\tl{\al}|\de_N|\right |,~~(\tl M_4^0)^2=|M|^2(1+|\de_N|)^2+|m_BM|\left
|1+\tl{\al}|\de_N|\right |
\la{M0s}
\eeq

The coupling of $\tl{n}^0$ states with the fermions emerges from the $F$-term of the superpotential $l^TY_{\nu }Nh_u$.
Following the transformations, indicated above, we will have:
$$
(l^TY_{\nu }Nh_u)_F\to \tl h_ul^TY_{\nu }\tl N=e^{-i\tl{\om}_2/2}\tl h_ul^T Y_{\nu }U_N\l \rho_ue^{i(\tl{\om}_2-\tl{\om}_1)/2},\rho_d\r
\tl{n}^0~,
$$
\beq
{\rm with}~~~
\rho_u=\fr{1}{\sq{2}}\l  \begin{array}{cc} 1&i\\0&0 \end{array}\r ~,~~~
\rho_d=\fr{1}{\sq{2}}\l  \begin{array}{cc} 0&0\\1&i \end{array}\r ~.
\la{YF0}
\eeq
Diagonalizing the matrix (\ref{Mtln}) by the transformation
\beq
V_{\tl n}^TM_{\tl n}^2V_{\tl n}=(M_{\tl n}^{Diag})^2,~~~~\tl{n}^0=V_{\tl n}\tl n ,
\la{diag-tiln0}
\eeq
the fermion coupling with the scalar  $\tl n$ mass eigenstates will be
\beq
\tl h_ul^TY_F\tl n~~~~~~{\rm with}~~~Y_F=Y_{\nu }\tl V^0 V_{\tl n}~,
~~~\tl V^0=U_N\l \rho_ue^{-i\tl{\om}_1/2},\rho_de^{-i\tl{\om}_2/2}\r ~.
\la{YF}
\eeq
The coupling with the slepton $\tl l$ is derived from the interaction term $h_u\tl l^T\l Y_{\nu }M_N^*\tl N^*-A_{\nu }\tl N\r $.
Going from  $\tl N$ to the $\tl n$ states, one obtains:
\beq
h_u\tl l^TY_B\tl n~~~~{\rm with}~~~Y_B=(Y_{\nu }M_N^*\tl V^{0*}-A_{\nu }\tl V^0)V_{\tl n}~.
\la{YB}
\eeq
For given values of $M, m_B$ and $m_A$, with help of Eqs. (\ref{Mtln}), (\ref{YF}) and (\ref{YB}), we will have coupling matrices $Y_F$,
$Y_B$ and all other quantities needed for calculation of the baryon asymmetry
created via the decays of the $\tl n_{1,2,3,4}$ states.

\subsection{Calculating $\fr{\tl n_b}{s}$ - Asymmetry Via $\tl{n}$ Decays}

 Due to the SUSY breaking terms, the masses of RHS's  differ from their fermionic partners' masses. For each mass-eigenstate RHS's
 $\tl{n}_{i=1,2,3,4}$ we have one of the masses $\tl{M}_{i=1,2,3,4}$ respectively.  With
the SUSY $M_S$ scale $\fr{M_S}{M}\stackrel{<}{_\sim }1/3$, the states $\tl n_i$ remain nearly degenerate and
for the resonant $\tl{n}$-decays the resummed effective amplitude technique \cite{Pilaftsis:1997jf} will be applied. Effective amplitudes
for the real $\tl{n}_i$ decay, say into the lepton $l_{\al }$ ($\al=1,2,3$) and antilepton $\ov{l}_{\al }$ respectively are given by
\cite{Pilaftsis:1997jf}
\beq
\hat{S}_{\al i}=S_{\al i}-\sum_jS_{\al j}\fr{\Pi_{ji}(\tl{M}_i)(1-\de_{ij})}{\tl{M}_i^2-\tl{M}_j^2+\Pi_{jj}(\tl{M}_i)}~,~~
\hat{\ov{S}}_{\al i}=S_{\al i}^*-\sum_jS_{\al j}^*\fr{\Pi_{ji}(\tl{M}_i)(1-\de_{ij})}{\tl{M}_i^2-\tl{M}_j^2+\Pi_{jj}(\tl{M}_i)}~,
\la{eff-amps}
\eeq
where $S_{\al i}$ is a tree level amplitude and $\Pi_{ij}$ is a two point Green function's (polarization operator of $\tl{n}_i-\tl{n}_j$)
absorptive part. The CP asymmetry is then given by
\beq
\ep_i^{sc}=\fr{\sum_{\al }\l |\hat{S}_{\al i}|^2-|\hat{\ov{S}}_{\al i}|^2\r}{\sum_{\al }\l |\hat{S}_{\al i}|^2+|\hat{\ov{S}}_{\al i}|^2\r }~.
\la{asym-byS}
\eeq

With $Y_F$ and $Y_B$ given by Eqs. (\ref{YF}) and (\ref{YB}) we can calculate  polarization diagram's (with external legs $\tl{n}_i$ and
$\tl{n}_j$) absorptive part $\Pi_{ij}$. These at 1-loop level are given by:
\beq
\Pi_{ij}(p)=\fr{ip^2}{8\pi}\!\l \!1-\fr{M_S^2}{p^2}\r^{\!\!2}\!\!\l Y_F^\dag Y_F+Y_F^T Y_F^*\r_{ij}\!+\fr{i}{8\pi}
\l \!s_{\beta}^2+c_{\beta}^2(1-\fr{M_S^2}{p^2}) \r \!\l Y_B^\dag Y_B+Y_B^TY_B^*\r_{ij} ~,
\la{abs-part}
\eeq
where $p$ denotes external momentum  in the diagram and upon evaluation  of (\ref{asym-byS}), for
 $\Pi $ one should use (\ref{abs-part}) with $p=\tl{M}_i$. In (\ref{abs-part}), taking into account the SUSY masses $M_S$ of all non SM states, we are using the refined expression for the $\Pi_{ij}$.

In an unbroken SUSY limit, neglecting finite temperature effects ($T\to 0$),
the $\tl{N}$ decay does not produce lepton asymmetry due to the  following reason. The decays of $\tl{N}$ in the fermion and scalar channels
are respectivelly
 $\tl{N}\to l\tl{h}_u$ and $\tl{N}\to \tl{l}^*h_u^*$. Since the rates of these processes are the same due to SUSY  (at $T=0$),
 the lepton asymmetries created from these decays cancel each other.
 With  $T\neq 0$, the cancellation does not take place  and one has
\beq
\tl{\ep }_i=\ep_i(\tl{n}_i\to l\tl{h}_u)\De_{BF}~,
\la{ep-non-zero}
\eeq
with a temperature dependent factor $\De_{BF}$ given in \cite{DAmbrosio:2003nfv}.\footnote{
This expression is valid with  alignment $A_{\nu }=m_AY_{\nu }$, which we are assuming to be true at the GUT scale and thus Eq. (\ref{ep-non-zero})
can be well applicable to our estimates.}
Therefore, we just need to compute $\ep_i(\tl{n}_i\to l\tl{h}_u)$, which is the asymmetry created by $\tl{n}_i$ decays in two fermions.
Thus, in (\ref{eff-amps}) we take $S_{\al i}=(Y_F)_{\al i}$ and calculate $\ep_i(\tl{n}_i\to l\tl{h}_u)$ with (\ref{asym-byS}).
The baryon asymmetry created from the lepton asymmetry due to $\tl{n}$ decays is given by:
\beq
\fr{\tl{n}_b}{s}\simeq -8.46\cdot 10^{-4}\sum_{i=1}^4\fr{\tl{\ep }_i}{\De_{BF}}\eta_i=
-8.46\cdot 10^{-4}\sum_{i=1}^4\ep_i(\tl{n}_i\to l\tl{h}_u)\eta_i~,
\la{sc-asym}
\eeq
where an effective number of degrees of freedom (including two RHN superfields) $g_*=228.75$ was used.
$\eta_i$ are efficiency factors which depend on $\tl{m}_i\simeq \fr{(v\sin \bt )^2}{M}2(Y_F^\dag Y_F)_{ii}$, and account for temperature effects once integration of the Boltzmann equations is performed \cite{DAmbrosio:2003nfv}.

Calculating the contribution $\fr{\De n_b}{s}=\fr{\tl n_b}{s}$ to the baryon asymmetry from the RHS decays, we have examined
various values of pairs $(m_A, m_B)$ in the range of $100$~GeV - few TeV. As it turned out, the ratio $\fr{\tl n_b}{n_b^f}$
is always suppressed($<3.4\cdot 10^{-2}$). The results for each neutrino scenario, we have considered in this paper, for
one specific choice of $(m_A, m_B)$, are given in Table \ref{scal-asym-for-textures} (see its caption for more information).
The ranges for $\fr{\tl n_b}{s}$ are due to the fact that for each scenario we have considered different values of $\tan \bt , M$ and $M_S$.
%
%
%
%
\begin{table}
\vs{0.1cm}
\hs{2.68cm}
 $$\begin{array}{|c|c|}
\hline
\vs{-0.3cm}
&\\
{\rm Neutrino ~Model}  & ~~~10^{11}\!\tm \!\fr{\tl n_b}{s}~~~~\\
\hline
{\rm Texture} ~P_1, ~{\rm NH}, ~{\rm data ~of~ tab.}~ \ref{tab01} & 0.23 - 0.28 \\
\hline
{\rm Texture} ~P_2, ~{\rm NH},~{\rm data ~of~ tab.}~ \ref{tab02} & 0.16 - 0.23\\
\hline
{\rm Texture} ~P_3, ~{\rm NH},~{\rm data ~of~ tab.}~  \ref{tab03} & \sim 0.1 \\
\hline
{\rm Texture} ~P_3, ~{\rm IH},~{\rm data ~of~ tab.}~  \ref{tab03} &0.07 - 0.09 \\
\hline
{\rm Texture} ~P_4, ~{\rm NH},~{\rm data ~of~ tab.}~  \ref{tab04}  & 0.07 - 0.08 \\
\hline
{\rm Texture ~A'}, ~{\rm NH},~{\rm data ~of~ Eqs.}~  (\ref{choice-Apr}), (\ref{mi-phases-Apr}) &0.05 - 0.07  \\
\hline
{\rm Texture} ~{\rm B_1}',~{\rm IH},~{\rm data ~of~ Eqs.}~  ( \ref{m3delroB1IH}), ~(\ref{mi-phasesB1IH}) & 0.04 - 0.049 \\
\hline
{\rm Texture} ~{\rm B_1}',~{\rm NH},~{\rm data ~of~ Eqs.}~  ( \ref{m3delroB1NH})-(\ref{btphi}) & \simeq 0 \\
\hline
{\rm Texture} ~{\rm B_2}',~{\rm IH},~{\rm data ~of~ Eqs.}~  (\ref{m3-delro-choice1}), (\ref{mi-phases}) & 0.042 - 0.05 \\
\hline
{\rm Texture} ~{\rm B_2}',~{\rm NH},~{\rm data ~of~ Eq.}~  (\ref{Bpr-NH}) &  \approx 1.4\times 10^{-4}\\
\hline
\end{array}$$
\vs{-0.5cm}
\caption{Values of $\fr{\De n_b}{s}=\fr{\tl n_b}{s}$ - contributions to the Baryon asymmetry via decays of the right handed sneutrinos
 for $(m_A, m_B)=(100i, 500)$~GeV and for various neutrino textures. Asymmetries are calculated with those values of $a_i$
and $b_j$ Yukawas that give $\l \fr{n_b}{s}\!\r_{\rm max}$. (For the latter see sections \ref{Resonant} and \ref{improvement}.) }
 \vs{-0.3cm}
 \label{scal-asym-for-textures}
\end{table}
Upon the calculations,
with obtained values
of $\tl{m}_i$, according to Ref. \cite{DAmbrosio:2003nfv} we picked up the corresponding values of $\eta_i$ and
used them in  (\ref{sc-asym}). While giving the results of the net baryon asymmetry, for each case (see sections \ref{Resonant} and \ref{improvement}), we have
included corresponding contributions from $\fr{\tl n_b}{s}$ as well. As we see from the results of Tab. \ref{scal-asym-for-textures},
the $\fr{\tl n_b}{s}$ is suppressed/subleading for all cases.
We have also witnessed  (by varying the phases of $m_{A, B}$) that the complexities of $m_A$ and $m_B$ practically do not change the results.
This happens because
the $m_A$ in the $Y_B$ coupling matrix appears in front of the $Y_{\nu }$ [see Eq. (\ref{YB})], which is strongly suppressed.
Irrelevance of the $m_B$'s phase can be seen from the structure of (\ref{Mtln}).
  Suppression of $\fr{\tl{n}_b}{s}$ will always happen for the value of
$|m_B|$ in the range of $100$ GeV - few TeV, because the mass degeneracy of $\tl{n}_i$ states is lifted in such a way that resonant
enhancement
of $\fr{\tl{n}_b}{s}$ is not realized. (Unlike  the case of soft leptogenesis \cite{DAmbrosio:2003nfv} which requires
$|m_B|\stackrel{<}{_\sim }10$~MeV. Without special arrangement, such suppressed values of $|m_B|$ seem unnatural and we have not considered
them  within our studies.)

\bibliographystyle{unsrt}

\end{document}